%% file: arxiv.tex
\RequirePackage{snapshot}

\documentclass[11pt]{lmcs}
    \input{nd-lmcs}

\title%
[Recurrence extraction and denotational semantics with recursive defintions]
{Recurrence extraction and denotational semantics with recursive function
definitions}
\author{N. Danner}

\usepackage{import}
\import{./}{packages}

\import{./}{macros}

\usepackage[colorlinks,allcolors=blue]{hyperref}
\newcommand{\doi}[1]{doi: \href{https://dx.doi.org/#1}{\url{#1}}}

\begin{document}
\maketitle

\import{./}{abstract}

\import{./}{main}

\bibliographystyle{abbrvnat}
\bibliography{../bib/refs}

\end{document}

%% file: nd-lmcs.tex
\makeatletter
\def\endfront@text{}
\makeatother

%% file: packages.tex
\usepackage{bussproofs}
	\input{nd-busscfg}

	\input{nd-typing}

\usepackage{float}
	\input{nd-float}

\usepackage[framemethod=tikz]{mdframed}
\usepackage{natbib}
\usepackage{suffix}
\usepackage{stmaryrd}
\usepackage{url}
\usepackage{wrapfig}
\usepackage{xcolor}

%% file: nd-busscfg.tex
\EnableBpAbbreviations
\def\doRule#1{\ifx#1\relax\else\RightLabel{#1}\fi}
\newcommand{\ndAXC}[2][\relax]{\AXC{$\mathstrut$}\doRule{#1}\UIC{#2}}
\newcommand{\ndUIC}[2][\relax]{\doRule{#1}\UIC{#2}}
\newcommand{\ndBIC}[2][\relax]{\doRule{#1}\BIC{#2}}
\newcommand{\ndTIC}[2][\relax]{\doRule{#1}\TIC{#2}}

%% file: nd-typing.tex
\let\proves\vdash

\def\oftype{\mathbin:}
\def\tmoftype#1#2{{#1}\oftype{#2}}
\newcommand{\typepf}[3][{}]{\proves_{#1}\tmoftype{#2}{#3}}

\newcommand{\typejudge}[4][{}]{{#2}\typepf[{#1}]{#3}{#4}}

%% file: nd-float.tex
  \makeatletter
  \newcommand{\fs@singlerule}{%
    \fs@plain%
    \def\@fs@mid{\vskip1ex\hrule\vskip1ex}
  }
  \makeatother
  \floatstyle{singlerule}
  \restylefloat{figure}
  \restylefloat{table}
  \newfloat{program}{t}{pgm}
  \floatname{program}{Program}

%% file: macros.tex

\newenvironment{callout}{%
  \begin{mdframed}[backgroundcolor=gray,roundcorner=8pt]
}{
  \end{mdframed}
}

%
{\begin{wrapfigure}{R}{#1}\begin{callout}}%
{\end{callout}\end{wrapfigure}}


\newcommand{\syncat}[1]{\underline{\smash{\mathrm{#1}}}}
\def\makesyncat#1{\expandafter\xdef\csname syn#1\endcsname{{\noexpand\syncat{#1}}}}
\foreach \x in {Type,Var,Term,Env} {\makesyncat{\x}}

\makeatletter

\newcommand{\NDinfruledef}[2]{\define@key{NDinf}{#1}[(#2)]{(#2)}}

\newcommand{\NDinfrule}[1]{\hbox{\setkeys{NDinf}{#1}}}

\NDinfruledef{refl}{refl}
\NDinfruledef{trans}{trans}
\NDinfruledef{eq-leq}{eq-leq}
\NDinfruledef{antisym}{antisym}
\NDinfruledef{mon}{mon}
\NDinfruledef{eq}{eq}
\NDinfruledef{zero}{zero}
\NDinfruledef{idl}{$+$-idl}
\NDinfruledef{idr}{$+$-idr}
\NDinfruledef{assoc}{$+$-assoc}
\NDinfruledef{val-cost}{val-cost}
\NDinfruledef{val-pot}{val-pot}
\NDinfruledef{bind-val}{bind-val}
\NDinfruledef{bind-cost}{bind-cost}
\NDinfruledef{bind-pot}{bind-pot}
\NDinfruledef{incr-cost}{incr-cost}
\NDinfruledef{incr-pot}{incr-pot}
\NDinfruledef{beta-to}{$\beta_\rightarrow$}
\NDinfruledef{beta-plus}{$\beta_{\mathord+}$}
\NDinfruledef{beta-times}{$\beta_{\mathord\times}$}
\NDinfruledef{beta-delta}{$\beta_\delta$}
\NDinfruledef{beta-fold}{$\beta_{\cfoldkw}$}
\NDinfruledef{beta-fix}{$\beta_{\cfixkw}$}

\newcommand{\infrulelbl}[1]{%
  \hypertarget{#1}{\NDinfrule{#1}}
}

\newcommand{\infruleref}[1]{%
  \hyperlink{#1}{\NDinfrule{#1}}}

\makeatother

\newcommand{\cross}{\times}

\newcommand{\bmax}{\vee}

\newcommand{\bigszlub}{\bigvee}
\newcommand{\bigszglb}{\bigwedge}
\newcommand{\floor}[1]{\lfloor#1\rfloor}
\newcommand{\ceil}[1]{\lceil#1\rceil}
\newcommand{\Implies}{\Rightarrow}
\newcommand{\Iff}{\Leftrightarrow}
\renewcommand{\And}{\wedge}

\newcommand{\comp}{\circ}
\newcommand{\disjunion}{\sqcup}
\newcommand{\bigintersection}{\bigcap}
\newcommand{\dom}{\mathop{\mathrm{dom}}\nolimits}

\def\setseparator{\mid}
\newcommand{\set}[2][\relax]{
  \ifx#1\relax
    \{#2\}
  \else
    \ifx#1\left
      \left\{#2\right\}
    \else
      \csname #1l\endcsname\{#2\csname #1r\endcsname\}
    \fi
  \fi
}
\newcommand{\setst}[3][\relax]{\set[#1]{#2\setseparator#3}}
\newcommand{\setidx}[3][\relax]{\set[#1]{#2}_{{#3}}}

\newenvironment{proofcases}{%
  \begin{description}%
}{\end{description}}

\newcommand{\proofcaselabel}[1]%
  {\hspace\labelsep{\textsc{Case: #1.}}\enspace}
\renewenvironment{proofcases}%
    {\begin{list}{}%
      {%
        
        \setlength{\labelwidth}{0in}
        \setlength{\leftmargin}{0in}
        \setlength{\itemsep}{2ex}
      }}%
    {\end{list}}

\newcommand{\skeyw}[1]{{\normalfont\texttt{#1}}}
\def\makeskeyw#1{\expandafter\xdef\csname s#1\endcsname{{\noexpand\skeyw{#1}}}}
\WithSuffix\def\makeskeyw*#1{\expandafter\xdef\csname s#1kw\endcsname{{\noexpand\skeyw{#1}}}}

\newcommand{\sint}{\skeyw{int}}
\newcommand{\sunit}{\skeyw{unit}}

\newcommand{\ssum}[2]{{#1}+{#2}}
\newcommand{\sprod}[2]{{#1}\cross {#2}}
\newcommand{\sarr}[2]{{#1}\to {#2}}
\newcommand{\slist}[1]{{#1}\,\skeyw{list}}

\newcommand{\slfp}[2]{\mu{#1}.{#2}}

\newcommand{\sconskw}{\skeyw{cons}}
\newcommand{\sempkw}{\skeyw{emp}}
\newcommand{\snodekw}{\skeyw{node}}
\newcommand{\sfunkw}{\skeyw{fun}}
\newcommand{\scasekw}{\skeyw{case}}
\newcommand{\sofkw}{\skeyw{of}}

\newcommand{\sifkw}{\skeyw{if}}
\newcommand{\sthenkw}{\skeyw{then}}
\newcommand{\selsekw}{\skeyw{else}}
\newcommand{\sfoldkw}{\skeyw{fold}}
\newcommand{\sunfoldkw}{\skeyw{unfold}}
\newcommand{\sletkw}{\skeyw{let}}
\newcommand{\sinkw}{\skeyw{in}}
\newcommand{\swild}{\skeyw{\underline{~}}}

\newcommand{\striv}{(\,)}

\newcommand{\sif}[3]{\sifkw\,#1\,\sthenkw\,#2\,\selsekw\,#3}
\WithSuffix\newcommand\sif*[3]{%
  \begin{aligned}[t]
  &\sifkw\,#1\,\sthenkw \\
  &\quad #2 \\
  &\selsekw \\
  &\quad #3
  \end{aligned}}
\WithSuffix\newcommand\sif^[3]{%
  \begin{aligned}[t]
  &\sifkw\,#1\,\sthenkw\,#2 \\
  &\selsekw\,#3
  \end{aligned}
}

\newcommand{\sinj}[2]{\sinkw_{#1}\,#2}
\WithSuffix\newcommand\sinj*[2]{\sinkw_{#1}(#2)}
\newcommand{\scase}[5]{%
  \scasekw\,#1\,\sofkw\,\sinj 0 {#2}\Rightarrow#3\mid\sinj 1 {#4}\Rightarrow#5}
\WithSuffix\newcommand\scase*[5]{%
  \scasekw\,#1\,\sofkw\,
  \begin{aligned}[t]
  &\phantom{{}\mid{}}\sinj 0 {#2}\Rightarrow#3 \\
  &\mid\sinj 1 {#4}\Rightarrow#5
  \end{aligned}}

\newcommand{\spair}[2]{(#1, #2)}

\newcommand{\sfun}[4][\relax]{\sfunkw\,#2\,#3\ifx#1\relax\else : #1\fi = #4}
\WithSuffix\newcommand\sfun*[4][\relax]{%
  \begin{aligned}[t]
  \sfunkw\,&#2\,#3\ifx#1\relax\else : #1\fi = \\
  &#4
  \end{aligned}}

\newcommand{\sapp}[2]{#1\,#2}
\newcommand{\snil}{\skeyw{nil}}
\newcommand{\scons}[2]{\sconskw(#1, #2)}
\newcommand{\scaselist}[5]{\scasekw\,#1\,\sofkw\,\snil\Rightarrow#2\mid\scons{#3}{#4}\Rightarrow#5}
\WithSuffix\newcommand\scaselist*[5]{%
  \scasekw\,#1\,\sofkw\,
  \begin{aligned}[t]
  &\phantom{{}\mid{}}\snil\Rightarrow#2 \\
  &\mid\scons{#3}{#4}\Rightarrow#5
  \end{aligned}}
\newcommand{\semp}{\sempkw}
\newcommand{\snode}[3]{\snodekw(#1, #2, #3)}
\newcommand{\scasetree}[6]{%
  \scasekw\,#1\,\sofkw\,\sempkw\Rightarrow#2\mid\snode{#3}{#4}{#5}\Rightarrow#6}
\WithSuffix\newcommand\scasetree*[6]{%
  \scasekw\,#1\,\sofkw\,
  \begin{aligned}[t]
  &\phantom{{}\mid{}}\semp\Rightarrow#2 \\
  &\mid\snode{#3}{#4}{#5}\Rightarrow#6
  \end{aligned}}
\newcommand{\sfold}[2][\relax]{\sfoldkw_{\ifx\#1\relax\else#1\fi}\,#2}
\newcommand{\sunfold}[2][\relax]{\sunfoldkw_{\ifx#1\relax\else#1\fi}\,#2}
\newcommand{\slet}[3]{\sletkw\,#1 = #2\,\sinkw\,#3}
\WithSuffix\newcommand\slet*[3]{%
  \begin{aligned}[t]
  &\sletkw\,#1 = #2 \\
  &\sinkw\,#3
  \end{aligned}}
\newcommand{\stick}[1]{{}^\checkmark#1}
\WithSuffix\newcommand\stick*[1]{\stick{(#1)}}


\foreach \x in {empty,isEmpty,enqueue,dequeue} 
  {\makeskeyw{\x}}


\foreach \x in {some,none} {\makeskeyw*{\x}}
\newcommand{\ssome}[1]{\ssomekw\,#1}
\WithSuffix\newcommand\ssome*[1]{\ssomekw(#1)}

\foreach \x in {abstype,is,with} {\makeskeyw*{\x}}
\newcommand{\sabstype}[3]{%
  \begin{aligned}[t]
  \sabstypekw\,&#1\,\siskw\,#2\,\swithkw \\
  #3
  \end{aligned}
}
\WithSuffix\newcommand\sabstype*[3]{%
  \sabstypekw\,#1\,\siskw\,#2\,\swithkw\,#3
}
\WithSuffix\newcommand\sabstypein*[4]{%
  \sabstypekw\,#1\,\siskw\,#2\,\swithkw\,#3\,\sinkw\,#4
}

\newcommand{\sctx}{\gamma}


\newcommand{\ckeyw}[1]{\mathsf{#1}}
\def\makeckeyw#1{\expandafter\xdef\csname c#1\endcsname{{\noexpand\ckeyw{#1}}}}
\WithSuffix\def\makeckeyw*#1{\expandafter\xdef\csname c#1kw\endcsname{{\noexpand\ckeyw{#1}}}}
\newcommand{\ccpy}[1]{\ckeyw{C}\,#1}
\WithSuffix\newcommand\ccpy*[1]{\ckeyw{C}(#1)}
\newcommand{\C}{\mathbb{C}}
\newcommand{\cint}{\ckeyw{int}}
\newcommand{\cunit}{\ckeyw{unit}}
\newcommand{\cbool}{\ckeyw{bool}}
\newcommand{\csum}[2]{{#1}+{#2}}
\newcommand{\cprod}[2]{{#1}\cross {#2}}
\newcommand{\carr}[2]{{#1}\to {#2}}
\newcommand{\clist}[1]{{#1}\,\ckeyw{list}}

\newcommand{\clfp}[2]{\mu{#1}.{#2}}

\newcommand{\cfixkw}{\ckeyw{fix}}

\newcommand{\cconskw}{\ckeyw{cons}}
\newcommand{\cempkw}{\ckeyw{emp}}
\newcommand{\cnodekw}{\ckeyw{node}}

\newcommand{\ccasekw}{\ckeyw{case}}
\newcommand{\cofkw}{\ckeyw{of}}

\newcommand{\cifkw}{\ckeyw{if}}
\newcommand{\cthenkw}{\ckeyw{then}}
\newcommand{\celsekw}{\ckeyw{else}}
\newcommand{\cfoldkw}{\ckeyw{fold}}
\newcommand{\cunfoldkw}{\ckeyw{unfold}}
\newcommand{\cletkw}{\ckeyw{let}}
\newcommand{\cinkw}{\ckeyw{in}}
\newcommand{\cwild}{\skeyw{\underline{~}}}

\makeckeyw*{val}
\makeckeyw*{bind}
\makeckeyw*{incr}

\newcommand{\ctriv}{(\,)}

\newcommand{\cif}[3]{\cifkw\,#1\,\cthenkw\,#2\,\celsekw\,#3}
\WithSuffix\newcommand\cif*[3]{%
  \begin{aligned}[t]
  &\cifkw\,#1\,\cthenkw \\
  &\quad #2 \\
  &\celsekw \\
  &\quad #3
  \end{aligned}}
\WithSuffix\newcommand\cif^[3]{%
  \begin{aligned}[t]
  &\cifkw\,#1\,\cthenkw\,#2 \\
  &\celsekw\,#3
  \end{aligned}
}

\newcommand{\cinj}[2]{\cinkw_{#1}\,#2}
\WithSuffix\newcommand\cinj*[2]{\cinkw_{#1}(#2)}
\newcommand{\ccase}[5]{%
  \ccasekw\,#1\,\cofkw\,\cinj 0 {#2}\Rightarrow#3\mid\cinj 1 {#4}\Rightarrow#5}
\WithSuffix\newcommand\ccase*[5]{%
  \ccasekw\,#1\,\cofkw\,
  \begin{aligned}[t]
  &\phantom{{}\mid{}}\cinj 0 {#2}\Rightarrow#3 \\
  &\mid\cinj 1 {#4}\Rightarrow#5
  \end{aligned}}
\newcommand{\ccaseset}[3]{%
  \ccasekw\,#1\,\cofkw\setidx{\cinj i {{#2}_i} \Rightarrow {#3}_i}{i=0,1}}

\newcommand{\ctuple}[1]{(#1)}
\newcommand{\cpair}[2]{\ctuple{#1, #2}}

\newcommand{\clam}[2]{\lambda #1.#2}
\newcommand{\cfix}[2]{\cfixkw\,#1.#2}
\WithSuffix\newcommand\cfix*[2]{%
\begin{aligned}[t]
\cfixkw\,&#1. \\
         &#2
\end{aligned}}
\newcommand{\capp}[2]{#1\,#2}
\newcommand{\cnil}{\ckeyw{nil}}
\newcommand{\ccons}[2]{\cconskw(#1, #2)}
\newcommand{\ccaselist}[5]{\ccasekw\,#1\,\cofkw\,\cnil\Rightarrow#2\mid\ccons{#3}{#4}\Rightarrow#5}
\WithSuffix\newcommand\ccaselist*[5]{%
  \ccasekw\,#1\,\cofkw\,
  \begin{aligned}[t]
  &\phantom{{}\mid{}}\cnil\Rightarrow#2 \\
  &\mid\ccons{#3}{#4}\Rightarrow#5
  \end{aligned}}
\WithSuffix\newcommand\ccaselist^[5]{%
  \begin{aligned}[t]
  &\ccasekw\,#1\,\cofkw\, \\
  &\phantom{{}\mid{}}\cnil\Rightarrow#2 \\
  &\mid\ccons{#3}{#4}\Rightarrow#5
  \end{aligned}}
\newcommand{\cemp}{\cempkw}
\newcommand{\cnode}[3]{\cnodekw(#1, #2, #3)}
\newcommand{\ccasetree}[6]{%
  \ccasekw\,#1\,\cofkw\,\cempkw\Rightarrow#2\mid\cnode{#3}{#4}{#5}\Rightarrow#6}
\WithSuffix\newcommand\ccasetree*[6]{%
  \ccasekw\,#1\,\cofkw\,
  \begin{aligned}[t]
  &\phantom{{}\mid{}}\cemp\Rightarrow#2 \\
  &\mid\cnode{#3}{#4}{#5}\Rightarrow#6
  \end{aligned}}
\WithSuffix\newcommand\ccasetree^[6]{%
  \begin{aligned}[t]
  &\ccasekw\,#1\,\cofkw \\
  &\phantom{{}\mid{}}\cemp\Rightarrow#2 \\
  &\mid\cnode{#3}{#4}{#5}\Rightarrow#6
  \end{aligned}}
\newcommand{\cfold}[2][\relax]{\cfoldkw_{\ifx#1\relax\else#1\fi}\,#2}
\newcommand{\cunfold}[2][\relax]{\cunfoldkw_{\ifx#1\relax\else#1\fi}\,#2}
\newcommand{\clet}[3]{\cletkw\,#1 = #2\,\cinkw\,#3}
\WithSuffix\newcommand\clet*[3]{%
  \begin{aligned}[t]
  &\cletkw\,#1 = #2 \\
  &\cinkw\,#3
  \end{aligned}}

\newcommand{\cval}[1]{\cvalkw\,#1}
\WithSuffix\newcommand\cval*[1]{\cvalkw(#1)}
\newcommand{\cbind}[3]{\cbindkw\,#1\gets#2\,\cinkw\,#3}
\WithSuffix\newcommand\cbind*[3]{%
  \begin{aligned}[t]
  &\cbindkw\,#1\gets#2 \\
  &\cinkw\,#3
  \end{aligned}
}
\newcommand{\cincr}[1]{\cincrkw\,#1}
\WithSuffix\newcommand\cincr*[1]{\cincrkw(#1)}
\newcommand{\ccost}[1]{#1_c}
\WithSuffix\newcommand\ccost*[1]{\ccost{\left(#1\right)}}
\newcommand{\cpot}[1]{#1_{\smash p}}
\WithSuffix\newcommand\cpot*[1]{\cpot{\left(#1\right)}}


\foreach \x in {empty,isEmpty,enqueue,dequeue} 
  {\expandafter\xdef\csname c\x\endcsname{{\noexpand\ckeyw{\x}}}}

\newcommand{\coption}[1]{{#1}\,\ckeyw{option}}
\WithSuffix\newcommand\coption*[1]{({#1})\,\ckeyw{option}}
\foreach \x in {some,none} {\makeckeyw*{\x}}
\newcommand{\csome}[1]{\csomekw\,#1}
\WithSuffix\newcommand\csome*[1]{\csomekw(#1)}

\foreach \x in {abstype,is,with} {\makeckeyw*{\x}}
\newcommand{\cabstype}[3]{%
  \begin{aligned}[t]
  \cabstypekw\,&#1\,\ciskw\,#2\,\cwithkw \\
  #3
  \end{aligned}
}
\WithSuffix\newcommand\cabstype*[3]{%
  \cabstypekw\,#1\,\ciskw\,#2\,\cwithkw\,#3
}
\WithSuffix\newcommand\cabstypein*[4]{%
  \cabstypekw\,#1\,\ciskw\,#2\,\cwithkw\,#3\,\cinkw\,#4
}

\newcommand{\eval}[3]{#1\downarrow^{#3}{#2}}
\newcommand{\oldeval}[4]{#1\downarrow^{#3}{#2}}
\WithSuffix\newcommand\eval*[3]{#1\downarrow^{#3}#2}
\newcommand{\ieval}[2]{#1\downarrow^{#2}{}}
\newcommand{\ievalold}[3]{#1\downarrow^{#2,#3}{}}
\newcommand{\red}[1][{}]{\rightarrow^{\phantom*}_{#1}}
\WithSuffix\newcommand\red*[1][{}]{\rightarrow^*_{#1}}

\newcommand{\subst}[3]{#1[#2/#3]}

\newcommand{\bdd}[1][\relax]{\preceq_{\ifx#1\relax\else#1\fi}}
\newcommand{\vbdd}[1][\relax]{\bdd[#1]^{\mathrm{val}}}
\newcommand{\bounded}[3][\relax]{{#2}\bdd[#1]{#3}}
\newcommand{\vbounded}[3][\relax]{{#2}\vbdd[#1]{#3}}
\newcommand{\oldbdd}[2][\relax]{\preceq_{\ifx#1\relax#2\else#1,#2\fi}}

\newcommand{\ctrans}[1]{\|#1\|}
\newcommand{\ptrans}[1]{\langle\!\langle#1\rangle\!\rangle}


\foreach \x in {add, to, withcost} {\makeckeyw*{\x}}
\newcommand{\withcost}[2]{{#1}\,\cwithcostkw\,{#2}}
\newcommand{\czcost}[1]{\withcost{#1}{0}}
\WithSuffix\newcommand\czcost*[1]{\withcost{#1}{0}}
\newcommand{\plusc}[2]{\caddkw\,{#1}\,\ctokw\,{#2}}
\WithSuffix\newcommand\plusc*[2]{%
\begin{aligned}[t]
&\caddkw\,#1 \\
&\ctokw\,#2
\end{aligned}
}


\newcommand\model[1]{\mathbf{#1}}

\newcommand{\N}{\model{N}}
\newcommand{\Ninfty}{\N^\infty}

\newcommand{\llambda}{\lambda\hskip-.45em\lambda}

\newcommand{\den}[2][\relax]{%
  \left\llbracket#2\right\rrbracket\ifx\#1\relax\else#1\fi}
\newcommand{\env}[1]{\{#1\}}
\newcommand{\envbind}[2]{#1\mapsto #2}
\newcommand{\envext}[3]{#1\env{\envbind{#2}{#3}}}

\newcommand{\cctx}{\gamma}
\newcommand{\szeq}{=}
\newcommand{\szleq}{\leq}

\newcommand{\szgeq}{\geq}
\newcommand{\szjudge}[4]{\typejudge{#1}{#2\szleq#3}{#4}}
\newcommand{\infleq}{\sqsubseteq}
\newcommand{\elimhole}[1]{[#1]}
\newcommand{\elimctx}[1]{\mathcal C\elimhole{#1}}

\newcommand{\smfix}{\mathit{fix}}

\newcommand\union\cup
\newcommand{\bigunion}{\bigcup}

\newcommand{\powerset}{\mathcal P}

\newcommand{\smfst}{\mathit{fst}}
\newcommand{\smsnd}{\mathit{snd}}
\newcommand{\sminj}[1]{\mathit{in}_{#1}}

\newcommand{\smcsize}{\mathit{csize}}
\newcommand{\smsucc}{\mathit{succ}}

\def\makesmfn#1{\expandafter\xdef\csname sm#1\endcsname{{\noexpand\mathit{#1}}}}
\foreach \x in {abs,conc} {\makesmfn{\x}}

%% file: abstract.tex
\begin{abstract}

With one exception, our previous work on recurrence extraction and
denotational semantics has focused on a source language that supports
inductive types and structural recursion.  The exception handles general
recursion via an initial translation into call-by-push-value.  In this note
we give an extraction function from a language with general recursive
function definitions and recursive types directly to a PCF-like recurrence
language.  We prove the main soundness result (that the syntactic
recurrences in fact bound the operational cost) without the use of a logical
relation, thereby significantly simplying the proof compared to our previous
work (at the cost of placing more demands on the models of the recurrence
language).  We then define two models of the recurrence language, one for
analyzing merge sort, and another for analyzing quick sort, as case studies
to understand model defintiions for justifying the extracted recurrences.

\end{abstract}


%% file: main.tex
We will not go over the fundamentals of the project of which this note is
part, and direct the reader to \citet{danner-licata:jfp22} for a
thorough discussion.  The key point is that we use denotational
semantics to give a foundation for the process most of us teach for
analyzing program cost:  extracting a recurrence that describes the cost of
the program in terms of the size of the input.  We factor this into two
phases:  a syntactic extraction from the program source language into a
language of recurrences, and then an interpretation of the extracted
recurrence into a model.  Most of our previous work
\citep{danner-et-al:icfp15,cutler-et-al:icpf20,danner-licata:jfp22}
uses a source language with inductive types and a structural recursion
operator in order to focus on the main ideas.  Supporting general recursive
function definitions raises (at least) two issues.  The first is that the
result that connects the bound described by the extracted syntactic
recurrence must take into account non-termination, and a necessary condition
is that if the extracted recurrence describes a finite bound on the cost,
then the original program must terminate.  The second is that since the
recurrence language is now some flavor of PCF, its models must be defined
with two orders simultaneously, one of which satisfies the size order
axioms, and the other of which corresponds to an information order, and
those two orders must interact.  \citet{kavvos-et-al:popl2020} provide one
approach to handling these issues.  On the syntactic side, the source
language from which recurrences are extracted is actually call-by-push-value
(CBPV) \citep{levy:cbpv}.  Standard translations of call-by-value and
call-by-name into CBPV yield recurrence extraction functions for the
intended languages.  However, the current author finds some of the details
of the analysis of the syntactic recurrence language somewhat unsatisfying.
For example, there is a precise operational semantics that must be respected
in order to discuss non-termination, and that seems somewhat counter to
viewing the language as a substitute for ordinary mathematical reasoning.
On the semantic side, the models are not investigated in detail, and it
behooves us to ensure that models such as those in
\citet{danner-licata:jfp22} carry over into a framework that supports
general recursion.

The main point of this note is to ensure that general recursion can be
handled smoothly, and that we can in fact define models that justify
informally extracted recurrences, something we have asserted can be done in
our prior papers.  This is not a deep dive, and there are certainly details
that have exposed lacunae in my background knowledge that need to be filled
so that they can be addressed.  I only occasionally point out relevant
results from the literature, and certainly haven't developed a thorough
review of related work.

\section{Source and recurrence languages}

\subsection{The source language}

\begin{wrapfigure}{R}{.45\linewidth}
\begin{callout}
In fact, we never really do anything with recursive types; see
Section~\ref{sec:dont-know}.
\end{callout}
\end{wrapfigure}

The source language is given in
Figures~\ref{fig:src-lang}-\ref{fig:src-lang-eval-int}.  Type
variables~$\alpha$ are only used for recursive types~$\slfp\alpha\sigma$;
typing only assigns closed types to terms.  In the presence of recursive
types, recursive functions can be defined without an explicit term former
(see, e.g., \citet[Ch.~20.1]{pierce:TAPL}).  We include one anyway, because
we will ultimately be concerned with models (of the recurrence language).
To keep the models somewhat simple, we will restrict our attention to
recursive types in which the body is a polynomial over~$\alpha$, and such
types are not sufficient to implement a fixpoint combinator.  However, it
turns out that proving the main bounding theorem is not difficult in the
presence of recursive types, so we include them because we can.

The evaluation judgment $\eval e v c$ is annotated with a cost~$c$ that
indicates resource usage, and which is incremented by a programmer-supplied
tick operator $\stick e$.  Following
\cite{hoffmann-et-al:toplas12:multivariate-amortized}, we define both a
standard big-step operational semantics along with an incomplete operational
semantics that derives judgments of the form~$\ieval e c$, indicating
that~$e$ has an incomplete or intermediate evaluation with cost~$c$.
Ignoring the annotations, using incomplete evaluation is a way of
incorporating some of the value of small-step semantics into a big-step
setting.  It lets us talk about ``reducing'' an expression without requiring
that we end up with a value.  This is useful if we want to be able to refer
to some part of the evaluation, but we do not care about the result when the
evaluation has not yet resulted in a value.  That is the setting we are in
for this work, because if we have not gotten a value yet, then we only care
about the cost of the evaluation so far.  Once we get a value, we care about
what it is.

\begin{figure}
\[
\begin{array}{rrl}
\sigma, \tau
  &::=& \alpha\
        \mid \sunit 
        \mid \ssum\sigma\sigma
        \mid \sprod\sigma\sigma 
        \mid \sarr\sigma\sigma 
        \mid \slfp\alpha\sigma
\\ \\
e
  &::=& x 
        \mid \striv 
        \mid \sinj 0 e \mid \sinj 1 e \mid \scase e x e x e
        \mid \spair e e \mid \slet {\spair x y} e e \\
  &\mid& \sfun f x e \mid \sapp e e 
         \mid \sfold e \mid \sunfold e \\
  &\mid& \stick e
\\ \\
v
  &::=& \striv \mid \sinj i v \mid \spair v v \mid \sfun f x e \mid \sfold v
\end{array}
\]
\caption{Source language grammar.}
\label{fig:src-lang}
\end{figure}

\begin{figure}
\begin{gather*}
\ndAXC{$\typejudge{\sctx,x\oftype\sigma}{x}{\sigma}$}
\DisplayProof
\qquad
\ndAXC{$\typejudge\sctx\striv\sunit$}
\DisplayProof
\\ \\
  \AXC{$\typejudge\sctx e {\sigma_i}$}
\ndUIC{$\typejudge\sctx{\sinj i e}{\ssum{\sigma_0}{\sigma_1}}$}
\DisplayProof
\qquad
  \AXC{$\typejudge\sctx e {\ssum{\sigma_0}{\sigma_1}}$}
  \AXC{$\typejudge{\sctx,x\oftype\sigma_0}{e_0}{\sigma}$}
  \AXC{$\typejudge{\sctx,x\oftype\sigma_1}{e_1}{\sigma}$}
\ndTIC{$\typejudge\sctx{\scase e x {e_0} x {e_1}}{\sigma}$}
\DisplayProof
\\ \\
  \AXC{$\typejudge\sctx{e_0}{\sigma_0}$}
  \AXC{$\typejudge\sctx{e_1}{\sigma_1}$}
\ndBIC{$\typejudge\sctx{\spair {e_0} {e_1}}{\sprod{\sigma_0}{\sigma_1}}$}
\DisplayProof
\qquad
  \AXC{$\typejudge\sctx{e'}{\sprod{\sigma_0}{\sigma_1}}$}
  \AXC{$\typejudge{\sctx,x_0\oftype\sigma_0,x_1\oftype\sigma_1}
                  {e}
                  {\sigma}$}
\ndBIC{$\typejudge\sctx{\slet {\spair {x_0} {x_1}} {e'} e}{\sigma}$}
\DisplayProof
\\ \\
  \AXC{$\typejudge{\sctx,f\oftype\sarr\sigma\tau,x\oftype\sigma}{e}{\tau}$}
\ndUIC{$\typejudge\sctx{\sfun f x e}{\sarr\sigma\tau}$}
\DisplayProof
\qquad
  \AXC{$\typejudge\sctx{e_0}{\sarr\sigma\tau}$}
  \AXC{$\typejudge\sctx{e_1}{\sigma}$}
\ndBIC{$\typejudge\sctx{\sapp{e_0}{e_1}}{\tau}$}
\DisplayProof
\\ \\
  \AXC{$\typejudge\sctx{e}{\subst\sigma{\slfp\alpha\sigma}{\alpha}}$}
\ndUIC{$\typejudge\sctx{\sfold e}{\slfp \alpha \sigma}$}
\DisplayProof
\qquad
  \AXC{$\typejudge\sctx e {\slfp \alpha \sigma}$}
\ndUIC{
  $\typejudge\sctx{\sunfold e}{\subst\sigma{\slfp\alpha\sigma}{\alpha}}$
}
\DisplayProof
\end{gather*}
\caption{Source language typing.}
\label{fig:src-lang-typing}
\end{figure}

\begin{figure}
\begin{gather*}
\ndAXC{$\eval\striv\striv 0$}
\DisplayProof
\qquad
\\ \\
  \AXC{$\eval e v c$}
\ndUIC{$\eval{\sinj i e}{\sinj i v}{c}$}
\DisplayProof
\qquad
  \AXC{$\eval{e}{\sinj i v}{c}$}
  \AXC{$\eval{\subst{e_i}{v}{x}}{v_i}{c_i}$}
\ndBIC{$\eval{\scase e x {e_0} x {e_1}}{v_i}{c+c_i}$}
\DisplayProof
\\ \\
  \AXC{$\setidx{\eval{e_i}{v_i}{c_i}}{i=0,1}$}
\ndUIC{$\eval{\spair{e_0}{e_1}}{\spair{v_0}{v_1}}{c_0+c_1}$}
\DisplayProof
\qquad
  \AXC{$\eval{e'}{\spair{v_0}{v_1}}{c'}$}
  \AXC{$\eval{\subst e {v_0,v_1} {x_0,x_1}} v {c}$}
\ndBIC{$\eval{\slet {\spair{x_0}{x_1}} {e'} {e}} v {c'+c} $}
\DisplayProof
\\ \\
\ndAXC{$\oldeval{\sfun f x e}{\sfun f x e} 0 0$}
\DisplayProof
\\
  \AXC{$\eval{e_0}{\sfun f x {e_0'}}{c_0}$}
  \AXC{$\eval{e_1}{v_1}{c_1}$}
  \AXC{$\eval{\subst{e_0'}{\sfun f x {e_0'},v_1}{f,x}}{v}{c}$}
\ndTIC{$\eval{\sapp{e_0}{e_1}}{v}{c_0+c_1+c}$}
\DisplayProof
\\ \\
  \AXC{$\eval{e}{v}{c}$}
\ndUIC{$\eval{\sfold e}{\sfold v}{c}$}
\DisplayProof
\qquad
  \AXC{$\eval{e}{\sfold v}{c}$}
\ndUIC{$\eval{\sunfold e}{v}{c}$}
\DisplayProof
\\ \\
  \AXC{$\eval{e}{v}{c}$}
\ndUIC{$\eval{\stick e}{v}{c+1}$}
\DisplayProof
\end{gather*}
\caption{Source language evaluation.}
\label{fig:src-lang-eval}
\end{figure}

\begin{figure}
\begin{gather*}
\ndAXC{$\ieval e 0$}
\DisplayProof
\qquad
\ndAXC{$\ieval\striv 0$}
\DisplayProof
\\ \\
  \AXC{$\ieval e c$}
\ndUIC{$\ieval{\sinj i e}{c}$}
\DisplayProof
\\
  \AXC{$\ieval{e}{c}$}
\ndUIC{$\ieval{\scase e  x {e_0} {x} {e_1}}{c}$}
\DisplayProof
\qquad
  \AXC{$\eval{e}{\sinj i v}{c}$}
  \AXC{$\ieval{\subst {e_i} {v} {x}}{c'}$}
\ndBIC{$\ieval{\scase e x {e_0} x {e_1}} {c+c'}$}
\DisplayProof
\\ \\
  \AXC{$\ieval{e_0}{c_0}$}
\ndUIC{$\ieval{\spair{e_0}{e_1}}{c_0}$}
\DisplayProof
\qquad
  \AXC{$\eval{e_0}{v_0}{c_0}$}
  \AXC{$\ieval{e_1}{c_1}$}
\ndBIC{$\ieval{\spair{e_0}{e_1}}{c_0+c_1}$}
\DisplayProof
\\
  \AXC{$\ieval {e'} {c'} $}
\ndUIC{$\ieval{\slet {\spair{x_0}{x_1}} {e'} {e}} {c'} $}
\DisplayProof
\qquad
  \AXC{$\eval {e'} {\spair{v_0}{v_1}} {c'} $}
  \AXC{$\ieval{\subst e {v_0,v_1} {x_0,x_1}} {c} $}
\ndBIC{$\ieval{\slet {\spair{x_0}{x_1}} {e'} {e}} {c'+c} $}
\DisplayProof
\\ \\
\ndAXC{$\ieval{\sfun f x e}{0}$}
\DisplayProof
\\
  \AXC{$\ieval{e_0}{c_0}$}
\ndUIC{$\ieval{\sapp{e_0}{e_1}}{c_0}$}
\DisplayProof
\qquad
  \AXC{$\oldeval{e_0}{\sfun f x {e_0'}}{c_0}{k_0}$}
  \AXC{$\ieval{e_1}{c_1}$}
\ndBIC{$\ieval{\sapp{e_0}{e_1}}{c_0+c_1}$}
\DisplayProof
\\
  \AXC{$\eval{e_0}{\sfun f x {e_0'}}{c_0}$}
  \AXC{$\eval{e_1}{v_1}{c_1}$}
  \AXC{$\ieval{\subst{e_0'}{\sfun f x {e_0'},v_1}{f,x}}{c}$}
\ndTIC{$\ieval{\sapp{e_0}{e_1}}{c_0+c_1+c}$}
\DisplayProof
\\ \\
  \AXC{$\ieval e c$}
\ndUIC{$\ieval{\sfold e} c$}
\DisplayProof
\qquad
  \AXC{$\ieval e c$}
\ndUIC{$\ieval{\sunfold e} c$}
\DisplayProof
\\ \\
  \AXC{$\ieval{e}{c}$}
\ndUIC{$\ieval{\stick e}{c+1}$}
\DisplayProof
\end{gather*}
\caption{Source language incomplete evaluation.}
\label{fig:src-lang-eval-int}
\end{figure}

We first make some expected observations about evaluation.

\begin{prop}
\label{prop:value-eval}
For any value~$v$, $\oldeval v v 0 0$.
\end{prop}

\begin{prop}
For every closed~$e$, there is at most one~$c$ and~$v$ such that
$\eval e v c$.
\end{prop}

\begin{prop}
If $\eval e v c$ and $\ieval e {c'} $, then $c'\leq c$.
\end{prop}

\subsection{The recurrence language}

\begin{wrapfigure}{R}{.65\linewidth}
\begin{callout}
The monadic type constructor for complexities is a change from previous
work, where a complexity type was (transparently) $\cprod\C\cdot$.  I would
like a nicer name for the $\cbindkw$ construct for the monad, but I haven't
been able to think of one.  In fact, I'm not really even sure $\cvalkw$ is
the right name for the return.
\end{callout}
\end{wrapfigure}

The recurrence language is the same as the source language, except that
$\sfun f x e$ is replaced with $\cfix x e$, a cost type~$\C$ is added,
and a monadic type constructor~$\ccpy\cdot$ is used for complexity
expressions.  The grammar is given in Figure~\ref{fig:rec-lang} and typing
rules in Figure~\ref{fig:rec-lang-typing}.
As expected, $\ccpy\cdot$ has an $\cincrkw$ operation
corresponding to increasing cost by~$1$.  
Informally a complexity (value of type~$\ccpy\sigma$) consists of a cost
and potential, and so we have projections $\ccost*\cdot$ and $\cpot*\cdot$
from~$\ccpy\sigma$ to the complexity type~$\C$ and~$\sigma$.  This makes
$\ccpy\sigma$ look a lot like $\C\cross\sigma$, but because $\ccpy\sigma$
can hide additional ``state'' about the cost that is not reflected in the
cost projection, the first projection on~$\C\cross\sigma$ cannot be
\emph{uniquely} factored through it.   For example, $\ccpy\sigma$ could be
interpreted as $(\N\cross\N)\cross\sigma$ with
$\ccost{(a, b), x} = a+b$.  The first projection on $\C\cross\sigma$ could
then be factored as either $(a, x) \mapsto ((a, 0), x) \mapsto a$ or
$(a, x) \mapsto ((0, a), x) \mapsto a$.
$\cvalkw$ is the return operation of the
monad, and its intended interpretation of $\cval e$
is a $0$-cost complexity from~$e$.
The binding operation binds a tuple of complexities and provides their
potentials to the body; the intended semantics is that the cost of the
complexities is added onto that of the body.  When the tuple has length~$1$,
we drop the parentheses.

\begin{wrapfigure}{R}{.45\linewidth}
\begin{callout}
In practice $\ccpy\sigma$ only arises when $\sigma$ does not have the form
$\ccpy{\sigma'}$---i.e., $\sigma$ is a potential type.  Is there any value
in having different syntactic classes for potential and complexity types?
\end{callout}
\end{wrapfigure}

\begin{figure}
\[
\begin{array}{rrl}
\sigma, \tau
  &::=& \alpha
        \mid \C 
        \mid \cunit 
        \mid \csum\sigma\sigma 
        \mid \cprod\sigma\sigma 
        \mid \carr\sigma\sigma 
        \mid \clfp \alpha \sigma
        \mid \ccpy\sigma
\\ \\
e
  &::=&  x \mid 0 \mid 1 \mid e + e \\
  &\mid& \ctriv 
         \mid \cinj 0 e \mid \cinj 1 e \mid \ccase e x e x e
         \mid \cpair e e \mid \clet {\cpair x y} e e \\
  &\mid& \clam x e \mid \capp e e \mid \cfix x e
         \mid \cfold e \mid \cunfold e \\
  &\mid& \cval e \mid 
         \cbind{\ctuple{x_0,\dots,x_{n-1}}} 
               {\ctuple{e_0,\dots,e_{n-1}}} 
               e \mid 
         \cincr e \mid \ccost e \mid \cpot e
\end{array}
\]
\caption{Recurrence language grammar.}
\label{fig:rec-lang}
\end{figure}

\begin{figure}
\begin{gather*}
\ndAXC{$\typejudge{\cctx,x\oftype\sigma}{x}{\sigma}$}
\DisplayProof
\qquad
\ndAXC{$\typejudge\cctx\ctriv\cunit$}
\DisplayProof
\\ \\
\ndAXC{$\typejudge\cctx 0 \C$}
\DisplayProof
\qquad
\ndAXC{$\typejudge\cctx 1 \C$}
\DisplayProof
\qquad
  \AXC{$\typejudge\cctx{e_0}{\C}$}
  \AXC{$\typejudge\cctx{e_1}{\C}$}
\ndBIC{$\typejudge\cctx{e_0+e_1}{\C}$}
\DisplayProof
\\ \\
  \AXC{$\typejudge\cctx e {\sigma_i}$}
\ndUIC{$\typejudge\cctx{\cinj i e}{\csum{\sigma_0}{\sigma_1}}$}
\DisplayProof
\qquad
  \AXC{$\typejudge\cctx e {\csum{\sigma_0}{\sigma_1}}$}
  \AXC{$\setidx{\typejudge{\cctx,x\oftype\sigma_i}{e_i}{\sigma}}{i=0,1}$}
\ndBIC{$\typejudge\cctx{\ccase e x {e_0} x {e_1}}{\sigma}$}
\DisplayProof
\\ \\
  \AXC{$\typejudge\cctx{e_0}{\sigma_0}$}
  \AXC{$\typejudge\cctx{e_1}{\sigma_1}$}
\ndBIC{$\typejudge\cctx{\cpair {e_0} {e_1}}{\cprod{\sigma_0}{\sigma_1}}$}
\DisplayProof
\\
  \AXC{$\typejudge\cctx{e'}{\cprod{\sigma_0}{\sigma_1}}$}
  \AXC{$\typejudge{\cctx,x_0\oftype\sigma_0,x_1\oftype\sigma_1}
                  {e}
                  {\sigma}$}
\ndBIC{$\typejudge\cctx{\clet {\cpair {x_0} {x_1}} {e'} e}{\sigma}$}
\DisplayProof
\\ \\
  \AXC{$\typejudge{\cctx,x\oftype\sigma}{e}{\tau}$}
\ndUIC{$\typejudge\cctx{\clam x e}{\carr\sigma\tau}$}
\DisplayProof
\qquad
  \AXC{$\typejudge\cctx{e_0}{\carr\sigma\tau}$}
  \AXC{$\typejudge\cctx{e_1}{\sigma}$}
\ndBIC{$\typejudge\cctx{\capp{e_0}{e_1}}{\tau}$}
\DisplayProof
\\ \\
  \AXC{$\typejudge\cctx{e}{\subst\sigma{\clfp \alpha\sigma}{\alpha}}$}
\ndUIC{$\typejudge\cctx{\cfold e}{\clfp \alpha \sigma}$}
\DisplayProof
\qquad
  \AXC{$\typejudge\cctx{e}{\clfp \alpha \sigma}$}
\ndUIC{$\typejudge\cctx{\cunfold e}{\subst\sigma{\clfp \alpha\sigma}{\alpha}}$}
\DisplayProof
\\ \\
  \AXC{$\typejudge{\cctx,x\oftype\sigma}{e}{\sigma}$}
\ndUIC{$\typejudge\cctx{\cfix x e}{\sigma}$}
\DisplayProof
\\ \\
  \AXC{$\typejudge\cctx e \sigma$}
\ndUIC{$\typejudge\cctx{\cval e}{\ccpy\sigma}$}
\DisplayProof
\qquad
  \AXC{$\typejudge\cctx {e_0} {\ccpy{\tau_0}}\dotsc
        \typejudge\cctx {e_{n-1}} {\ccpy{\tau_{n-1}}}$}
  \AXC{$\typejudge{\cctx,x_0\oftype\tau_0,\dots,x_{n-1}\oftype\tau_{n-1}}
                  {e}
                  {\ccpy\sigma}$}
\ndBIC{$\typejudge\cctx{\cbind{\ctuple{x_0,\dots,x_{n-1}}}
                              {\ctuple{e_0,\dots,e_{n-1}}} 
                              {e}}
                       {\ccpy\sigma}$}
\DisplayProof
\\
  \AXC{$\typejudge\cctx e {\ccpy\sigma}$}
\ndUIC{$\typejudge\cctx{\cincr e}{\ccpy\sigma}$}
\DisplayProof
\\
  \AXC{$\typejudge\cctx e {\ccpy\sigma}$}
\ndUIC{$\typejudge\cctx{\ccost e}{\C}$}
\DisplayProof
\qquad
  \AXC{$\typejudge\cctx e {\ccpy\sigma}$}
\ndUIC{$\typejudge\cctx{\cpot e}{\sigma}$}
\DisplayProof
\end{gather*}
\caption{Recurrence language typing.}
\label{fig:rec-lang-typing}
\end{figure}

Using the monadic type primarily just buys us some cleaner notation for the
recurrence extraction function.  It also lets us carve out exactly where we
want the features of negative products in terms of being able to project
exact cost and potential from complexity expressions while interpreting
products positively, which gives us finer-grained control of the semantics
(see the discussion of quick sort in Section~\ref{sec:quick-sort}.

The size (pre)order is defined in Figure~\ref{fig:size-order}.  Formally the
inference rules presuppose a type judgment, so that, for example,
\infruleref{refl} and \infruleref{trans} are really
\[
\AXC{$\typejudge\gamma e \sigma$}
\ndUIC{$\typejudge\gamma {e\szleq e} \sigma$}
\DisplayProof
\qquad
\AXC{$\typejudge\gamma{e\szleq e'}{\sigma}$}
\AXC{$\typejudge\gamma{e'\szleq e''}{\sigma}$}
\ndBIC{$\typejudge\gamma{e\szleq e''}{\sigma}$}
\DisplayProof
\]
and so forth.  $\elimctx{}$ ranges over contexts for which we need to ensure
monotonicity of $\szleq$.  We define $e = e'$ to mean $e\szleq e'$ and
$e'\szleq e$ and observe that $=$ also satisfies \infruleref{refl},
\infruleref{trans}, and~\infruleref{mon}.

\begin{figure}
\begin{gather*}
\begin{split}
\elimctx{} ::=
  &\phantom{{}\mid{}}     \elimhole{}
  \mid \elimctx{} + e \mid e + \elimctx{} 
  \mid \ccost*{\elimctx{}} \mid \cpot*{\elimctx{}} \\
  & \mid \cinj i {\elimctx{}} \mid \ccaseset{\elimctx{}}{x}{e}
    \mid \cpair{\elimctx{}}{e} \mid \cpair{e}{\elimctx{}}
    \mid \clet{\cpair{x_0}{x_1}}{\elimctx{}}{e} \\
  & \mid \capp{\elimctx{}}{e} \mid \capp{e}{\elimctx{}}
    \mid \cfold{\elimctx{}} \mid \cunfold{\elimctx{}}
\end{split}
\\
\RightLabel{\infrulelbl{refl}}
\ndAXC{$e\szleq e$}
\DisplayProof
\qquad
  \AXC{$e\szleq e'$}
  \AXC{$e'\szleq e''$}
\RightLabel{\infrulelbl{trans}}
\ndBIC{$e\szleq e''$}
\DisplayProof
\qquad
  \AXC{$e\szleq e'$}
\RightLabel{\infrulelbl{mon}}
\ndUIC{$\elimctx e\szleq \elimctx{e'}$}
\DisplayProof
\\
\RightLabel{\infrulelbl{zero}}
\ndAXC{$0\szleq e$}
\DisplayProof
\qquad
\RightLabel{\infrulelbl{idl}}
\ndAXC{$0 + e \szeq e$}
\DisplayProof
\qquad
\RightLabel{\infrulelbl{idr}}
\ndAXC{$e + 0 \szeq e$}
\DisplayProof
\RightLabel{\infrulelbl{assoc}}
\ndAXC{$(e + e') + e'' \szeq e + (e' + e'')$}
\DisplayProof
\\
\RightLabel{\infrulelbl{bind-val}}
\ndAXC{$\cbind p {\cval {e'}} {e} = \subst{e}{e'}{p}$}
\DisplayProof
\\
\RightLabel{\infrulelbl{val-cost}}
\ndAXC{$\ccost*{\cval e} \szeq 0$}
\DisplayProof
\qquad
\RightLabel{\infrulelbl{val-pot}}
\ndAXC{$\cpot*{\cval e} \szeq e$}
\DisplayProof
\\
\RightLabel{\infrulelbl{bind-cost}}
\ndAXC{$
\begin{aligned}
& \ccost*{e_0}+\dots+\ccost*{e_{n-1}} + 
  \ccost*{\subst e {\cpot*{e_0},\dots,\cpot*{e_{n-1}}} 
                   {x_0,\dots,x_{n-1}}} 
  \szeq \\
&\qquad
  \ccost*{\cbind{\ctuple{x_0,\dots,x_{n-1}}}
                {\ctuple{e_0,\dots,e_{n-1}}}
                {e}}
\end{aligned}
$}
\DisplayProof
\\
\RightLabel{\infrulelbl{bind-pot}}
\ndAXC{$\cpot*{\subst e {\cpot*{e_0},\dots,\cpot*{e_{n-1}}} 
                        {x_0,\dots,x_{n-1}}} 
        \szeq
        \cpot*{\cbind{\ctuple{x_0,\dots,x_{n-1}}}
                     {\ctuple{e_0,\dots,e_{n-1}}}
                     {e}}$}
\DisplayProof
\\
\RightLabel{\infrulelbl{incr-cost}}
\ndAXC{$\ccost e + 1 \szeq \ccost*{\cincr e}$}
\DisplayProof
\qquad
\RightLabel{\infrulelbl{incr-pot}}
\ndAXC{$\cpot e \szeq \cpot*{\cincr e}$}
\DisplayProof
\\ \\
\RightLabel{\infrulelbl{beta-plus}}
\ndAXC{$\subst{e_i}{e}{x}\szleq\ccase{\cinj i e} x {e_0} x {e_1}$}
\DisplayProof
\\
\RightLabel{\infrulelbl{beta-times}}
\ndAXC{$\subst e {e_0,e_1} {x_0,x_1}\szleq
        \clet {\cpair{x_0}{x_1}} {\cpair{e_0}{e_1}} e$}
\DisplayProof
\\
\RightLabel{\infrulelbl{beta-to}}
\ndAXC{$\subst e {e_1} x \szleq
        \capp{(\clam x e)}{e_1}$}
\DisplayProof
\\
\RightLabel{\infrulelbl{beta-fold}}
\ndAXC{$e\szleq \cunfold{(\cfold e)}$}
\DisplayProof
\\
\RightLabel{\infrulelbl{beta-fix}}
\ndAXC{$\subst e {\cfix x e} {x} \szleq \cfix x e$}
\DisplayProof
\end{gather*}
\caption{The size (pre)order.  Also see Section~\ref{sec:size-order-more}
and Figure~\ref{fig:size-order-more}.}
\label{fig:size-order}
\end{figure}

\subsection{Recurrence extraction and soundness}

Recurrence extraction is defined in Figure~\ref{fig:rec-extraction}.
We define an extraction for both values and arbitrary expressions in order
to prove an appropriate soundness theorem.

\begin{figure}
\begin{gather*}
\ctrans\sigma = \ccpy{\ptrans\sigma}
\\
\ptrans\sunit = \cunit 
\\
\ptrans{\ssum{\sigma_0}{\sigma_1}} =
  \cprod{\ptrans{\sigma_0}}{\ptrans{\sigma_1}}
\qquad
\ptrans{\sprod{\sigma_0}{\sigma_1}} =
  \cprod{\ptrans{\sigma_0}}{\ptrans{\sigma_1}}
\qquad
\ptrans{\sarr{\sigma}{\tau}} =
  \carr{\ptrans{\sigma}}{\ctrans{\tau}}
\\
\ptrans{\slfp\alpha\sigma} = \clfp\alpha{\ptrans\sigma}
\end{gather*}

\begin{align*}
\ctrans x &= \cval x \\
\ctrans \striv &= \cval \ctriv \\
\ctrans{\sinj i e}
  &= \cbind{p}{\ctrans e}{\cval*{\cinj i p}} \\
\ctrans{\scase e x {e_0} x {e_1}}
  &= \cbind p {\ctrans e} {\ccase p x {\ctrans{e_0}} x {\ctrans{e_1}}} \\
\ctrans {\spair{e_0}{e_1}}
  &= \cbind{\cpair{p_0}{p_1}} 
           {\cpair{\ctrans{e_0}}{\ctrans{e_1}}}
           {\cval{\cpair{p_0}{p_1}}} \\
\ctrans {\slet {\spair{x_0}{x_1}} {e'} {e}}
  &= \cbind p {\ctrans {e'}} {\clet {\cpair{x_0}{x_1}} p {\ctrans e}} \\
\ctrans {\sfun f x e}
  &= \cval* {\cfix f {(\clam x {\ctrans e})}} \\
\ctrans {\sapp{e_0}{e_1}}
  &= \cbind {\cpair{p_0}{p_1}} {\cpair{e_0}{e_1}} {\capp{p_0}{p_1}} \\
\ctrans{\sfold e}
  &= \cbind p {\ctrans e} {\cval*{\cfold p}} \\
\ctrans{\sunfold e}
  &= \cbind p {\ctrans e} {\cval*{\cunfold p}} \\
\ctrans {\stick e}
  &= \cincr{\ctrans e}
\\ \\
\ptrans\striv &= \striv \\
\ptrans{\sinj i v} &= \cinj i {\ptrans v} \\
\ptrans{\spair{v_0}{v_1}} &= \cpair{\ptrans{v_0}}{\ptrans{v_1}} \\
\ptrans{\sfun f x e} &= \cfix f {(\clam x {\ctrans e}} \\
\ptrans{\sfold v} &= \cfold{\ptrans v}
\end{align*}
\caption{Recurrence extraction.}
\label{fig:rec-extraction}
\end{figure}

\begin{lem}[Value extraction]
For all values~$v$, $\ctrans v = \cval\ptrans v$.
\end{lem}
\begin{proof}
The proof is by induction on~$v$; we'll just do a few cases.
\begin{proofcases}
\item[$\striv$]:  $\ctrans\striv = \cval\ctriv = \cval*{\ptrans\striv}$.

\item[$\spair{v_0}{v_1}$]
\begin{align*}
\ctrans{\spair{v_0}{v_1}}
  &= \cbind {p_0,p_1} {\ctrans{v_0}, \ctrans{v_1}} {\cval\cpair{p_0}{p_1}} \\
  &= \cbind {p_0,p_1} {\cval\ptrans{v_0}, \cval\ptrans{v_1}} {\cval\cpair{p_0}{p_1}} \\
  &= \cval\cpair{\ptrans{v_0}}{\ptrans{v_1}} \\
  &= \cval\ptrans{\spair{v_0}{v_1}}.
\end{align*}

\item[$\sfold v$]
\begin{align*}
\ctrans{\sfold v}
  &= \cbind p {\ctrans v} {\cfold p} \\
  &= \cbind p {\cval\ptrans v} {\cfold p} \\
  &= \cfold{\ptrans v} \\
  &= \ptrans{\sfold v}.
\end{align*}
\end{proofcases}
\end{proof}

\begin{lem}[Substitution]
\label{lem:extr-subst}
For all expressions~$e$ and values~$\vec v$,
$\ctrans{\subst e {\vec v} {\vec x}} = 
\subst{\ctrans e}{\ptrans{\vec v}}{\vec x}$, where we write
$\ptrans{v_0,\dots,v_{n-1}}$ for $\ptrans{v_0},\dots,\ptrans{v_{n-1}}$.
\end{lem}
\begin{proof}
The proof is by induction on~$e$; we'll just do a few cases.
\begin{proofcases}
\item[$x_i$]
\[
\ctrans{\subst{x_i}{\vec v}{\vec x}}
= \ctrans{v_i}
= \cval\ptrans{v_i}
= \subst{(\cval{x_i})}{\ptrans{\vec v}}{\vec x}
= \subst{\ctrans{x_i}}{\ptrans{\vec v}}{\vec x}.
\]

\item[$\slet{\spair y z} {e'} {e}$]
\begin{align*}
\ctrans{\subst{(\slet{\spair y z} {e'} {e})}{\vec v}{\vec x}}
  &= \ctrans{\slet
               {\spair y z} 
               {\subst{e'}{\vec v}{\vec x}}
               {\subst{e}{\vec v}{\vec x}}} \\
  &= \cbind p {\ctrans{\subst{e'}{\vec v}{\vec x}}}
            {\clet {\cpair y z} p {\ctrans{\subst e {\vec v}{\vec x}}}} \\
  &= \cbind p {\subst{\ctrans{e'}}{\ptrans{\vec v}}{\vec x}}
            {\clet {\cpair y z} p {\subst{\ctrans{e}} {\ptrans{\vec v}}{\vec x}}} \\
  &= \subst{(
       \cbind p {\ctrans{e'}} {\clet {\cpair y z} p {\ctrans{e}}}
     )}{\ptrans{\vec v}}{\vec x} \\
  &= \subst{\ctrans{\slet{\spair y z}{e'}{e}}}{\ptrans{\vec v}}{\vec x}.
\end{align*}
\end{proofcases}
\end{proof}


\begin{wrapfigure}{R}{.45\linewidth}
\begin{callout}
It seems like it would be more ``monadic'' to avoid the cost and potential
projections and formulate Soundness as
something like the following:  if $\eval e v c$, then
$\cincrkw^c(\cval\ptrans v)\szleq \ctrans e$, where
$\cincrkw^c\,e = \cincr*{\cdots\cincr e}$ ($c$ times).  But it isn't clear
what to say when we only have $\ieval e c$ and not $\eval e v c$.  Note that
this property is equivalent to part~(2) of the Soundness Theorem.
\end{callout}
\end{wrapfigure}

As one might expect, the Soundness theorem tells us that the cost projection
of an extracted recurrence bounds the actual cost, and the potential
projection bounds the value when there is one.  In the past we've used a
logical relation between source and recurrence language programs, but Dan
Licata observed that \citet{van-stone:thesis} proved a comparable result in
her thesis without one, so we should be able to do the same.

\begin{thm}[Soundness]
\label{thm:soundness-thm}
\label{thm:bounding-thm}
For all~$e$:
\begin{enumerate}
\item If $\ieval e c$, then $c\szleq \ccost{\ctrans e}$.
\item If $\eval e v c$, then $c\szleq\ccost{\ctrans e}$ and
$\ptrans v\szleq \cpot{\ctrans e}$.
\end{enumerate}
\end{thm}

\begin{proof}
We prove (1) and~(2) simultaneously by induction on the height of the
(incomplete) evaluation derivation, breaking into cases according to the
last line of the derivation.

\begin{proofcases}
\item[$
\ndAXC{$\eval {\striv}{\striv}{0}$}
\DisplayProof
$]
$0= 0$ and 
$\ptrans{\striv} = \ctriv = \cpot*{\cval{\ctriv}} = \cpot{\ctrans\striv}$.

\item[$
  \AXC{$\eval e v c$}
\ndUIC{$\eval{\sinj i e}{\sinj i v}{c}$}
\DisplayProof
$]
By the induction hypothesis, $c\leq\ccost{\ctrans e}$ and
$\ptrans v\szleq \cpot{\ctrans e}$.  Thus:
\[
\begin{aligned}[t]
\ccost{\ctrans{\sinj i e}}
  &\szeq\ccost*{\cbind p {\ctrans e} {\cval*{\cinj i p}}} \\
  &\szeq\ccost{\ctrans e} + \ccost*{\cval*{\cinj i {\cpot{\ctrans e}}}} 
  & & \infruleref{bind-cost} \\
  &\szgeq c + 0 & & \text{IH}, \infruleref{val-cost}, \infruleref{mon} \\
  &\szeq c & & \infruleref{idr}
\end{aligned}
\]
and
\[
\begin{aligned}[t]
\cpot{\ctrans{\sinj i e}}
  &\szeq\cpot*{\cbind p {\ctrans e} {\cval*{\cinj i p}}}  \\
  &\szeq\cpot*{\cval*{\cinj i {\cpot{\ctrans e}}}}
  & & \infruleref{bind-pot} \\
  &\szeq\cinj i {\cpot{\ctrans e}}
  & & \infruleref{val-pot} \\
  &\szgeq \cinj i {\ptrans v} & & \text{IH}, \infruleref{mon} \\
  &\szeq\ptrans{\sinj i v}
\end{aligned}
\]
Monotonicity uses the contexts $\elimctx{}+0$ and $\cinj i {\elimctx{}}$.

\item[$
  \AXC{$\ieval e c$}
\ndUIC{$\ieval{\sinj i e}{c}$}
\DisplayProof
$]
By the induction hypothesis, $c\szleq \ccost{\ctrans e}$.  The reasoning is
the same as in the cost analysis of the previous case.

\item[$
  \AXC{$\eval{e}{\sinj i v}{c}$}
  \AXC{$\eval{\subst{e_i}{v}{x}}{v_i}{c_i}$}
\ndBIC{$\eval{\scase e x {e_0} x {e_1}}{v_i}{c+c_i}$}
\DisplayProof
$]
By the induction hypothesis, $c\szleq \ccost{\ctrans e}$,
$\cinj i {\ptrans v} = \ptrans {\sinj i v}\szleq \cpot{\ctrans e}$,
$c_i\szleq \ccost{\ctrans{\subst{e_i}{v}{x}}}$, and
$\ptrans{v_i}\szleq \cpot{\ctrans{\subst{e_i}{v}{x}}}$.  Thus:
\begin{multline*}
\ccost{\ctrans{\scase e x {e_0} x {e_1}}} \\
\begin{aligned}
  &\szeq  \ccost*{\cbind p {\ctrans e} 
                         {\ccase p x {\ctrans{e_0}} x {\ctrans{e_1}}}} \\
  &\szeq  \ccost{\ctrans e} +
          \ccost*{\ccase
                    {\cpot{\ctrans e}}
                    {x} {\ctrans{e_0}}
                    {x} {\ctrans{e_1}}} \\
  &\szgeq c +
          \ccost*{\ccase
                    {\cinj i {\ptrans v}}
                    {x} {\ctrans{e_0}}
                    {x} {\ctrans{e_1}}}
  & & \text{IH}, \infruleref{mon} \\
  &\szgeq c + \ccost*{\subst{\ctrans{e_i}}{\ptrans v}{x}}
  & & \infruleref{beta-plus}, \infruleref{mon} \\
  &\szeq  c + \ccost{\ctrans{\subst{e_i}{v}{x}}}
  & & \text{Lemma~\ref{lem:extr-subst}} \\
  &\szgeq c + c_i
  & & \text{IH}, \infruleref{mon}
\end{aligned}
\end{multline*}
and
\begin{multline*}
\cpot{\ctrans{\scase e x {e_0} x {e_1}}} \\
\begin{aligned}
  &\szeq  \cpot*{\cbind p {\ctrans e} 
                         {\ccase p x {\ctrans{e_0}} x {\ctrans{e_1}}}} \\
  &\szeq  \cpot*{\ccase
                    {\cpot{\ctrans e}}
                    {x} {\ctrans{e_0}}
                    {x} {\ctrans{e_1}}} \\
  &\szgeq \cpot*{\ccase
                    {\cinj i {\ptrans v}}
                    {x} {\ctrans{e_0}}
                    {x} {\ctrans{e_1}}}
  & & \text{IH}, \infruleref{mon} \\
  &\szgeq \cpot*{\subst{\ctrans{e_i}}{\ptrans v}{x}}
  & & \infruleref{beta-plus}, \infruleref{mon} \\
  &\szeq  \cpot{\ctrans{\subst{e_i}{v}{x}}}
  & & \text{Lemma~\ref{lem:extr-subst}} \\
  &\szgeq \ptrans{v_i}
  & & \text{IH}.
\end{aligned}
\end{multline*}
Monotonicity uses the contexts 
$\ccase{\elimctx{}}{x}{\ctrans{e_0}}{x}{\ctrans{e1}}$,
$\ccost*{\elimctx{}}$, $\cpot*{\elimctx{}}$, and
$c+\elimctx{}$

\item[$
  \AXC{$\ieval{e}{c}$}
\ndUIC{$\ieval{\scase e  x {e_0} {x} {e_1}}{c}$}
\DisplayProof
$]
By the induction hypothesis, $c\leq \ccost{\ctrans e}$.  Thus:
\begin{align*}
\ccost{\ctrans{\scase e x {e_0} x {e_1}}}
  &\szeq  \ccost*{\cbind p {\ctrans e} 
                         {\ccase p x {\ctrans{e_0}} x {\ctrans{e_1}}}} \\
  &\szeq  \ccost{\ctrans e} +
          \ccost*{\ccase
                    {\cpot{\ctrans e}}
                    {x} {\ctrans{e_0}}
                    {x} {\ctrans{e_1}}} \\
  &\szgeq \ccost{\ctrans e} + 0 \\
  &\szgeq \ccost{\ctrans e}.
\end{align*}

\item[$
  \AXC{$\eval{e}{\sinj i v}{c}$}
  \AXC{$\ieval{\subst {e_i} {v} {x}}{c'}$}
\ndBIC{$\ieval{\scase e x {e_0} x {e_1}} {c+c'}$}
\DisplayProof
$]
The argument is the same as the cost argument for the complete evaluation.

\item[$
  \AXC{$\setidx{\eval{e_i}{v_i}{c_i}}{i=0,1}$}
\ndUIC{$\eval{\spair{e_0}{e_1}}{\spair{v_0}{v_1}}{c_0+c_1}$}
\DisplayProof
$]
By the induction hypothesis, $c_i\szleq\ccost{\ctrans{e_i}}$ and
$\ptrans{v_i}\szleq\cpot{\ctrans{e_i}}$.  Thus:
\begin{align*}
\ccost{\ctrans{\spair{e_0}{e_1}}}
  &\szeq  \ccost*{\cbind {p_0,p_1} {\ctrans{e_0}, \ctrans{e_1}}
                         {\cval{\cpair{p_0}{p_1}}}} \\
  &\szeq  \ccost{\ctrans{e_0}} + \ccost{\ctrans{e_1}} +
          \ccost*{\cval{\cpair{\cpot{\ctrans{e_0}}}{\cpot{\ctrans{e_1}}}}}
          \\
  &\szgeq c_0 + c_1 + 0 \\
  &\szeq  c_0 + c_1
\end{align*}
and
\begin{align*}
\cpot{\ctrans{\spair{e_0}{e_1}}}
  &\szeq  \cpot*{\cbind {p_0,p_1} {\ctrans{e_0}, \ctrans{e_1}}
                         {\cval{\cpair{p_0}{p_1}}}} \\
  &\szeq  \cpot*{\cval{\cpair{\cpot{\ctrans{e_0}}}{\cpot{\ctrans{e_1}}}}}
          \\
  &\szeq  \cpair{\cpot{\ctrans{e_0}}}{\cpot{\ctrans{e_1}}} \\
  &\szgeq \cpair{\ptrans{v_0}}{\ptrans{v_1}} \\
  &\szeq  \ptrans{\spair{v_0}{v_1}}.
\end{align*}

\item[$
  \AXC{$\ieval{e_0}{c_0}$}
\ndUIC{$\ieval{\spair{e_0}{e_1}}{c_0}$}
\DisplayProof
$]
By the induction hypothesis, $c_0\szleq \ccost{\ctrans{e_0}}$.  Thus:
\begin{align*}
\ccost{\ctrans{\spair{e_0}{e_1}}}
  &\szeq  \ccost*{\cbind {p_0,p_1} {\ctrans{e_0}, \ctrans{e_1}}
                         {\cval{\cpair{p_0}{p_1}}}} \\
  &\szeq  \ccost{\ctrans{e_0}} + \ccost{\ctrans{e_1}} +
          \ccost*{\cval{\cpair{\cpot{\ctrans{e_0}}}{\cpot{\ctrans{e_1}}}}}
          \\
  &\szgeq c_0 + 0 + 0 \\
  &\szeq  c_0.
\end{align*}

\item[$
  \AXC{$\eval{e_0}{v_0}{c_0}$}
  \AXC{$\ieval{e_1}{c_1}$}
\ndBIC{$\ieval{\spair{e_0}{e_1}}{c_0+c_1}$}
\DisplayProof
$]
By the induction hypothesis, $c_0\szleq \ccost{\ctrans{e_0}}$ and
$c_1\szleq\ccost{\ctrans{e_1}}$.  Thus:
\begin{align*}
\ccost{\ctrans{\spair{e_0}{e_1}}}
  &\szeq  \ccost*{\cbind {p_0,p_1} {\ctrans{e_0}, \ctrans{e_1}}
                         {\cval{\cpair{p_0}{p_1}}}} \\
  &\szeq  \ccost{\ctrans{e_0}} + \ccost{\ctrans{e_1}} +
          \ccost*{\cval{\cpair{\cpot{\ctrans{e_0}}}{\cpot{\ctrans{e_1}}}}}
          \\
  &\szgeq c_0 + c_1 + 0 \\
  &\szeq  c_0 + c_1.
\end{align*}

\item[$
  \AXC{$\eval{e'}{\spair{v_0}{v_1}}{c'}$}
  \AXC{$\eval{\subst e {v_0,v_1} {x_0,x_1}} v {c}$}
\ndBIC{$\eval{\slet {\spair{x_0}{x_1}} {e'} {e}} v {c'+c} $}
\DisplayProof
$]
By the induction hypothesis $c'\szleq \ccost{\ctrans{e'}}$,
$\cpair{\ptrans{v_0}}{\ptrans{v_1}} =
\ptrans{\spair{v_0}{v_1}}\szleq \cpot{\ctrans{e'}}$,
$c\szleq \ccost{\ctrans{\subst{e}{v_0,v_1}{x_0,x_1}}}$, and
$\ptrans v\szleq\cpot{\ctrans{\subst{e}{v_0,v_1}{x_0,x_1}}}$.  Thus:
\begin{align*}
\ccost{\ctrans{\slet{\spair{x_0}{x_1}}{e'}{e}}}
  &= \ccost*{
       \cbind p {\ctrans {e'}} {
         \clet {\cpair{x_0}{x_1}} p {\ctrans{e}}
       }
     } \\
  &= \ccost{\ctrans{e'}} +
     \ccost*{ \clet {\cpair{x_0}{x_1}} {\cpot{\ctrans{e'}}} {\ctrans{e}} } \\
  &\szgeq c' +
          \ccost*{
            \clet {\cpair{x_0}{x_1}} {\cpair{\ptrans{v_0}}{\ptrans{v_1}}}
                  {\ctrans{e}}
          }
  & & \infruleref{mon} \\
  & \szgeq c' +
           \ccost*{
             \subst{\ctrans e}{\ptrans{v_0},\ptrans{v_1}}{x_0,x_1}
           } & & \infruleref{beta-times},\infruleref{mon} \\
  &= c' + \ccost*{\ctrans{\subst e {v_0,v_1}{x_0,x_1}}}
     & & \text{Lemma~\ref{lem:extr-subst}} \\
  &\szgeq c' + c  & & \text{IH}, \infruleref{mon}
\end{align*}
and
\begin{align*}
\cpot{\ctrans{\slet{\spair{x_0}{x_1}}{e'}{e}}}
  &= \cpot*{
       \cbind p {\ctrans {e'}} {
         \clet {\cpair{x_0}{x_1}} p {\ctrans{e}}
       }
     } \\
  &= \cpot*{ \clet {\cpair{x_0}{x_1}} {\cpot{\ctrans{e'}}} {\ctrans{e}} } \\
  &\szgeq \cpot*{
            \clet {\cpair{x_0}{x_1}} {\cpair{\ptrans{v_0}}{\ptrans{v_1}}}
                  {\ctrans{e}}
          }
  & & \infruleref{mon} \\
  & \szgeq \cpot*{
             \subst{\ctrans e}{\ptrans{v_0},\ptrans{v_1}}{x_0,x_1}
           } & & \infruleref{beta-times},\infruleref{mon} \\
  &= \cpot*{\ctrans{\subst e {v_0,v_1}{x_0,x_1}}}
     & & \text{Lemma~\ref{lem:extr-subst}} \\
  &\szgeq \ptrans v  & & \text{IH}.
\end{align*}
Monotonicity uses the contexts
$\elimctx{}+e$, $e+\elimctx{}$, $\ccost*{\elimctx{}}$, $\cpot*{\elimctx{}}$,
and $\clet{\cpair{x_0}{x_1}}{\elimctx{}}{e}$.

\item[$
\ndAXC{$\eval{\sfun f x e}{\sfun f x e}{0}$}
\DisplayProof
$]
The cost bound is immediate, and
\begin{align*}
\cpot{\ctrans{\sfun f x e}}
  &=     \cpot*{\cval*{\cfix f {\clam x {\ctrans e}}}} \\
  &\szeq \cfix f {\clam x {\ctrans e}} \\
  &=     \ptrans{\sfun f x e}.
\end{align*}

\item[$
  \AXC{$\eval{e_0}{\sfun f x {e_0'}}{c_0}$}
  \AXC{$\eval{e_1}{v_1}{c_1}$}
  \AXC{$\eval{\subst{e_0'}{\sfun f x {e_0'},v_1}{f,x}}{v}{c}$}
\ndTIC{$\eval{\sapp{e_0}{e_1}}{v}{c_0+c_1+c}$}
\DisplayProof
$]
By the induction hypothesis,
\[
\begin{aligned}[t]
c_0 &\szleq \ccost{\ctrans{e_0}} \\
\cfix f {\clam x {\ctrans{e_0'}}} =
  \ptrans{\sfun f x {e_0'}} &\szleq \cpot{\ctrans{e_0}}
\end{aligned}
\qquad
\begin{aligned}[t]
c_1 &\szleq \ccost{\ctrans{e_1}} \\
\ptrans{v_1} &\szleq \cpot{\ctrans{e_1}}
\end{aligned}
\]
and
\begin{align*}
c&\szleq\ccost{\ctrans{\subst{e_0'}{\sfun f x {e_0'},v_1}{f,x}}}
  = \subst{\ccost{\ctrans{e_0'}}}
          {\cfix f{\clam x {\ctrans{e_0'}}}, \ptrans{v_1}}
          {f, x} \\
\ptrans{v} 
  &\szleq \cpot{\ctrans{\subst{e_0'}{\sfun f x {e_0'},v_1}{f,x}}}
  = \subst{\cpot{\ctrans{e_0'}}}
          {\cfix f{\clam x {\ctrans{e_0'}}}, \ptrans{v_1}}
          {f, x}.
\end{align*}
Thus:
\begin{align*}
\ccost{\ctrans{\sapp{e_0}{e_1}}}
  &\szeq \ccost*{\cbind{p_0,p_1}{\ctrans{e_0},\ctrans{e_1}}
                       {\capp{p_0}{p_1}}} \\
  &\szeq \ccost{\ctrans{e_0}} + \ccost{\ctrans{e_1}} +
         \ccost*{\capp{\cpot{\ctrans{e_0}}}{\cpot{\ctrans{e_1}}}} \\
  &\szgeq c_0 + c_1 +
          \ccost*{\capp{(\cfix f {\clam x {\ctrans{e_0'}}})}{\ptrans{v_1}}}
          \\
  &\szgeq c_0 + c_1 +
          \ccost*{\subst{\ctrans{e_0'}}
                        {\cfix f {\clam x {\ctrans{e_0'}}}, \ptrans{v_1}}
                        {f, x}} \\
  &\szgeq c_0 + c_1 + c
\end{align*}
and
\begin{align*}
\cpot{\ctrans{\sapp{e_0}{e_1}}}
  &\szeq \cpot*{\cbind{p_0,p_1}{\ctrans{e_0},\ctrans{e_1}}
                       {\capp{p_0}{p_1}}} \\
  &\szeq \cpot*{\capp{\cpot{\ctrans{e_0}}}{\cpot{\ctrans{e_1}}}} \\
  &\szgeq \cpot*{\capp{(\cfix f {\clam x {\ctrans{e_0'}}})}{\ptrans{v_1}}}
          \\
  &\szgeq \cpot*{\subst{\ctrans{e_0'}}
                        {\cfix f {\clam x {\ctrans{e_0'}}}, \ptrans{v_1}}
                        {f, x}} \\
  &\szgeq \ptrans{v}.
\end{align*}

\item[$
  \AXC{$\ieval{e_0}{c_0}$}
\ndUIC{$\ieval{\sapp{e_0}{e_1}}{c_0}$}
\DisplayProof
$]
By the induction hypothesis, $c_0\szleq\ccost{\ctrans{e_0}}$.  Thus:
\begin{align*}
\ccost{\ctrans{\sapp{e_0}{e_1}}}
  &\szeq \ccost*{\cbind{p_0,p_1}{\ctrans{e_0},\ctrans{e_1}}
                       {\capp{p_0}{p_1}}} \\
  &\szeq \ccost{\ctrans{e_0}} + \ccost{\ctrans{e_1}} +
         \ccost*{\capp{\cpot{\ctrans{e_0}}}{\cpot{\ctrans{e_1}}}} \\
  &\szgeq c_0 + 0 + 0 \\
  &\szgeq c_0.
\end{align*}

\item[$
  \AXC{$\oldeval{e_0}{\sfun f x {e_0'}}{c_0}{k_0}$}
  \AXC{$\ieval{e_1}{c_1}$}
\ndBIC{$\ieval{\sapp{e_0}{e_1}}{c_0+c_1}$}
\DisplayProof
$]
By the induction hypothesis, $c_0\szleq\ccost{\ctrans{e_0}}$ and
$c_1\szleq\ccost{\ctrans{e_1}}$.  Thus:
\begin{align*}
\ccost{\ctrans{\sapp{e_0}{e_1}}}
  &\szeq \ccost*{\cbind{p_0,p_1}{\ctrans{e_0},\ctrans{e_1}}
                       {\capp{p_0}{p_1}}} \\
  &\szeq \ccost{\ctrans{e_0}} + \ccost{\ctrans{e_1}} +
         \ccost*{\capp{\cpot{\ctrans{e_0}}}{\cpot{\ctrans{e_1}}}} \\
  &\szgeq c_0 + c_1 + 0 \\
  &\szgeq c_0 + c_1.
\end{align*}

\item[$
  \AXC{$\eval{e_0}{\sfun f x {e_0'}}{c_0}$}
  \AXC{$\eval{e_1}{v_1}{c_1}$}
  \AXC{$\ieval{\subst{e_0'}{\sfun f x {e_0'},v_1}{f,x}}{c}$}
\ndTIC{$\ieval{\sapp{e_0}{e_1}}{c_0+c_1+c}$}
\DisplayProof
$]
The argument is the same as the cost argument for the complete evaluation.

\item[$
  \AXC{$\eval{e}{v}{c}$}
\ndUIC{$\eval{\sfold e}{\sfold v}{c}$}
\DisplayProof
$]
By the induction hypothesis, $c\szleq \ccost{\ctrans e}$ and
$\ptrans v\szleq \cpot{\ctrans e}$.  Thus:
\[
\begin{aligned}[t]
\ccost{\ctrans{\cfold e}}
  &=      \ccost*{\cbind p {\ctrans e} {\cfold p}} \\
  &\szeq  \ccost{\ctrans e} + \ccost*{\cval*{\cfold{\cpot{\ctrans e}}}} \\
  &\szgeq c + 0 \\
  &\szeq  c
\end{aligned}
\qquad\text{and}\qquad
\begin{aligned}[t]
\cpot{\ctrans{\cfold e}}
  &=      \cpot*{\cbind p {\ctrans e} {\cfold p}} \\
  &\szeq  \cpot*{\cval*{\cfold{\cpot{\ctrans e}}}} \\
  &\szeq  \cfold{\cpot{\ctrans e}} \\
  &\szgeq \cfold{\ptrans v} \\
  &=      \ptrans{\sfold v}.
\end{aligned}
\]

\item[$
  \AXC{$\ieval e c$}
\ndUIC{$\ieval{\sfold e} c$}
\DisplayProof
$]
The argument is the same as the cost argument for the complete evaluation.

\item[$
  \AXC{$\eval{e}{\sfold v}{c}$}
\ndUIC{$\eval{\sunfold e}{v}{c}$}
\DisplayProof
$]
By the induction hypothesis, $c\szleq\ccost{\ctrans{e}}$ and
$\cfold{\ptrans v} = \ptrans{\sfold v}\szleq \cpot{\ctrans{e}}$.  Thus:
\begin{align*}
\ccost{\ctrans{\sunfold e}}
  &=      \ccost*{\cbind p {\ctrans e} {\cval*{\cunfold p}}} \\
  &\szgeq \ccost{\ctrans e} + \ccost*{\cval*{\cunfold{\cpot{\ctrans e}}}} \\
  &\szgeq c + 0 \\
  &\szeq  c
\end{align*}
and
\begin{align*}
\cpot{\ctrans{\sunfold e}}
  &=      \cpot*{\cbind p {\ctrans e} {\cval*{\cunfold p}}} \\
  &\szgeq \cpot*{\cval*{\cunfold{\cpot{\ctrans e}}}} \\
  &\szeq  \cunfold{\cpot{\ctrans e}} \\
  &\szgeq \cunfold{(\cfold{\ptrans v})} \\
  &\szgeq \ptrans v.
\end{align*}

\item[$
  \AXC{$\ieval e c$}
\ndUIC{$\ieval{\sunfold e} c$}
\DisplayProof
$]
The argument is the same as for the cost argument for the complete
evaluation.

\item[$
  \AXC{$\eval{e}{v}{c}$}
\ndUIC{$\eval{\stick e}{v}{c+1}$}
\DisplayProof
$]
By the induction hypothesis, $c\szleq\ccost{\ctrans e}$ and
$\ptrans v\szleq \cpot{\ctrans e}$.  Thus:
\[
\begin{aligned}[t]
\ccost{\ctrans{\stick e}}
  &=      \ccost*{\cincr{\ctrans e}} \\
  &\szeq  \ccost{\ctrans e} + 1 \\
  &\szgeq c + 1
\end{aligned}
\quad\text{and}\quad
\begin{aligned}[t]
\cpot{\ctrans{\stick e}}
  &=      \cpot*{\cincr{\ctrans e}} \\
  &\szeq  \cpot{\ctrans e} \\
  &\szgeq \ptrans v.
\end{aligned}
\]

\item[$
  \AXC{$\ieval{e}{c}$}
\ndUIC{$\ieval{\stick e}{c+1}$}
\DisplayProof
$]
The argument is the same as for the cost argument for the complete
evaluation.\qedhere
\end{proofcases}
\end{proof}

\subsection{Bounding relations}

Our previous work establishes a Soundness-like result using a logical
relation between the source and recurrence languages.  We can recover that
sort of relation as follows.

\begin{defi}[Bounding relations]\hfill
\begin{enumerate}
\item
Define $\bounded e E$ to mean that if $\ieval e c$, then $c\leq \ccost E$,
and if $\eval e v c$, then $c\leq \ccost E$ and $\vbounded v {\cpot E}$.

\item
Define $\vbounded v E$ to mean $\ptrans v \szleq E$.
\end{enumerate}
\end{defi}

So the Soundness Theorem can be restated as saying that for all~$e$,
$\bounded e {\ctrans e}$.  The reason we care about the bounding relations
is that they tell us that Soundness ``propagates down by application:''

\begin{prop}
If $\vbounded{\sfun f x e}{E}$ and $\vbounded{v'}{E'}$, then
$\bounded{\subst e {\sfun f x e, v'} {f, x}}{\capp E {E'}}$.
\end{prop}

\subsection{A strengthening of the recurrence language}
\label{sec:size-order-more}

Figure~\ref{fig:size-order} provides the minimum constraints on~$\szleq$
that are needed in order to prove Theorem~\ref{thm:soundness-thm}.  As a
reminder, these constraints state that $\szleq$ is a preorder, the
equivalence relation~$=$ induced by $e\szleq e' \szleq e$ behaves nicely with
respect to the monadic operations, and that the main term constructors are
the source of size abstractions.  But in practice our models
interpret~$\szleq$ as a partial order and thus~$=$ is ordinary equality.  So
that we can reduce clutter in our exposition, we add corresponding inference
rules to the recurrence language in Figure~\ref{fig:size-order-more}.  In
that figure, $\mathcal T$ ranges over all term formers in
Figure~\ref{fig:rec-lang}.

\begin{figure}
\begin{gather*}
\AXC{$e = e'$}
\RightLabel{\infrulelbl{eq-leq}}
\ndUIC{$e\szleq e'$}
\DisplayProof
\qquad
\AXC{$e\szleq e'$}
\AXC{$e'\szleq e$}
\RightLabel{\infrulelbl{antisym}}
\ndBIC{$e = e'$}
\DisplayProof
\qquad
\AXC{$e_0 = e_0'$}
\AXC{$e_1 = e_1'$}
\AXC{$\ldots$}
\RightLabel{\infrulelbl{eq}}
\ndTIC{$\mathcal T(e_0,e_1,\dotsc) = \mathcal T(e_0',e_1',\dotsc)$}
\DisplayProof
\end{gather*}
\caption{More size order inference rules.}
\label{fig:size-order-more}
\end{figure}

\subsection{Lists}
\label{sec:lists}

For our examples we will work with the standard inductive types of lists 
along with the standard syntactic sugar for constructors:
\[
\begin{aligned}[t]
\slist\sigma &= \slfp \alpha {\csum{\cunit}{\cprod{\sigma}{\alpha}}} \\
\snil        &= (\sfoldkw_{\slist\sigma}\comp\sinkw_0)\striv \\
\sconskw     &= \sfoldkw_{\slist\sigma}\comp\sinkw_1
\end{aligned}
\]
We also have comparable notation in the recurrence language and observe that
\begin{align*}
\ctrans\snil
  &= \cbind
       p
       {\ctrans{\sinj 0 {\striv}}}
       {\cval*{\cfold p}} \\
  &= \cbind
       p
       {
         \cbind
           {p'}
           {\ctrans\striv}
           {\cval*{\cinj 0 {p'}}}
       }
       {\cval*{\cfold p}} \\
  &= \cbind
       p
       {
         \cbind
           {p'}
           {\cval\ctriv}
           {\cval*{\cinj 0 {p'}}}
       }
       {\cval*{\cfold p}} \\
  &= \cval*{
       \cfold{(\cinj 0 \ctriv)}
     } \\
  &= \cval\cnil.
\end{align*}

Unfortunately, the corresponding computation for the extraction of 
$\sconskw$ expressions is not quite so straightforward, because we need to
perform simplifications in $\cbindkw$ expressions, and we can only perform
non-trivial simplifications underneath a cost- or potential projection.
Thus we must perform two separate computations:
\begin{multline*}
\ccost{\ctrans{\scons{e}{e'}}} \\
\begin{aligned}
  &= \ccost*{
       \cbind
         p
         {
           \cbind
             {p'}
             {
               \cbind
                 {\cpair{q}{q'}}
                 {\cpair{\ctrans{e}}{\ctrans{e'}}}
                 {\cval{\cpair{q}{q'}}}
             }
             {\cval*{\cinj 1 {p'}}}
         }
         {\cval*{\cfold p}}
     } \\
  &= \ccost*{
         {
           \cbind
             {p'}
             {
               \cbind
                 {\cpair{q}{q'}}
                 {\cpair{\ctrans{e}}{\ctrans{e'}}}
                 {\cval{\cpair{q}{q'}}}
             }
             {\cval*{\cinj 1 {p'}}}
         }
     } \\
  &= \ccost*{
             {
               \cbind
                 {\cpair{q}{q'}}
                 {\cpair{\ctrans{e}}{\ctrans{e'}}}
                 {\cval{\cpair{q}{q'}}}
             }
     } \\
  &= \ccost{\ctrans e} + \ccost{\ctrans{e'}}
\end{aligned}
\end{multline*}
and
\begin{multline*}
\cpot{\ctrans{\scons{e}{e'}}} \\
\begin{aligned}
  &= \cpot*{
       \cbind
         p
         {
           \cbind
             {p'}
             {
               \cbind
                 {\cpair{q}{q'}}
                 {\cpair{\ctrans{e}}{\ctrans{e'}}}
                 {\cval{\cpair{q}{q'}}}
             }
             {\cval*{\cinj 1 {p'}}}
         }
         {\cval*{\cfold p}}
     } \\
  &= \cfold{
       \cpot*
         {
           \cbind
             {p'}
             {
               \cbind
                 {\cpair{q}{q'}}
                 {\cpair{\ctrans{e}}{\ctrans{e'}}}
                 {\cval{\cpair{q}{q'}}}
             }
             {\cval*{\cinj 1 {p'}}}
         }
     } \\
  &= \cfold{(\cinj 1 {
       \cpot*{
         \cbind
           {\cpair{q}{q'}}
           {\cpair{\ctrans{e}}{\ctrans{e'}}}
           {\cval{\cpair{q}{q'}}}
       }}
     )} \\
  &= \cfold{(\cinj 1 {\cpair{\cpot{\ctrans{e}}}{\cpot{\ctrans{e'}}})}} \\
  &= \ccons{\cpot{\ctrans e}}{\cpot{\ctrans{e'}}}.
\end{aligned}
\end{multline*}

We define the expression
\[
\scaselist e {e_{\snil}} {x} {xs} {e_{\sconskw}}
\]
to be syntactic sugar for
\[
\scase 
  {\sunfold[\slist\sigma] e} 
  {\swild} 
  {e_\snil} 
  {z}
  {\slet{\spair{x}{xs}}{z}{e_{\sconskw}}}
\]
and have similar sugar in the recurrence language.
As expected, the recurrence extracted from one is written in terms of the
other, but as with $\sconskw$ expressions, only underneath cost and
potential projections.

\begin{multline*}
\ccost{\ctrans{\scaselist e {e_{\snil}} {x} {xs} {e_{\sconskw}}}} \\
\begin{aligned}
  &= \ccost{
       \ctrans{
         \scase
           {\sunfold[\slist\sigma] e} 
           {\swild} 
           {e_\snil} 
           {z}
           {\slet{\spair{x}{xs}}{z}{e_{\sconskw}}}
       }
     } \\
  &= \bigl(\cbind
       p
       {\ctrans{\sunfold e}}
       {
         \ccase*
           p
           {\cwild}
           {\ctrans{e_\snil}}
           {z}
           {
             \ctrans{\slet{\spair{x}{xs}}{z}{e_{\sconskw}}}
           \bigr)_c
           }
        } \\
  &= \bigl(\cbind*
       p
       {
         \cbind
           {p'}
           {\ctrans e}
           {\cval*{\cunfold{p'}}}
       }
       {
         \ccase*
           p
           {\cwild}
           {\ctrans{e_\snil}}
           {z}
           {
             \ctrans{\slet{\spair{x}{xs}}{z}{e_{\sconskw}}}
           \bigr)_c
           }
        } \\
  &= \ccost*{
       {
         \cbind
           {p'}
           {\ctrans e}
           {\cval*{\cunfold{p'}}}
       }
     } + \\
  &  \qquad
     \bigl(
         \ccase*
           {
             \cpot*{
               \cbind
                 {p'}
                 {\ctrans e}
                 {\cval*{\cunfold{p'}}}
             }
           }
           {\cwild}
           {\ctrans{e_\snil}}
           {z}
           {
             \ctrans{\slet{\spair{x}{xs}}{z}{e_{\sconskw}}}
           \bigr)_c
           }
     \\
  &= \ccost{\ctrans e} +
     \ccost*{
       {\cval*{\cunfold{\cpot{\ctrans e}}}}
     } + \\
  &  \qquad
     \bigl(
         \ccase*
           {\cpot*{\cval*{\cunfold{\cpot{\ctrans e}}}}}
           {\cwild}
           {\ctrans{e_\snil}}
           {z}
           {
             \ctrans{\slet{\spair{x}{xs}}{z}{e_{\sconskw}}}
           \bigr)_c
           }
     \\
  &= \ccost{\ctrans e} +
     \bigl(
         \ccase*
           {\cunfold{\cpot{\ctrans e}}}
           {\cwild}
           {\ctrans{e_\snil}}
           {z}
           {
             \ctrans{\slet{\spair{x}{xs}}{z}{e_{\sconskw}}}
           \bigr)_c
           }
     \\
  &= \ccost{\ctrans e} +
     \bigl(
         \ccase*
           {\cunfold{\cpot{\ctrans e}}}
           {\cwild}
           {\ctrans{e_\snil}}
           {z}
           {
             \cbind
               q
               {\cval z}
               {
                 \clet
                   {\cpair{x}{xs}}
                   {q}
                   {\ctrans{e_\sconskw}}
               }
           \bigr)_c
           }
     \\
  &= \ccost{\ctrans e} +
     \bigl(
         \ccase*
           {\cunfold{\cpot{\ctrans e}}}
           {\cwild}
           {\ctrans{e_\snil}}
           {z}
           {
                 \clet
                   {\cpair{x}{xs}}
                   {z}
                   {\ctrans{e_\sconskw}}
           \bigr)_c
           }
     \\
  &= \ccost{\ctrans e} +
     \ccost*{\ccaselist
       {\cpot{\ctrans e}}
       {\ctrans{e_{\snil}}}
       {x}{xs}
       {\ctrans{e_{\sconskw}}}
     }.
\end{aligned}
\end{multline*}
and a similar calculation shows that
\begin{multline*}
\cpot{\ctrans{\scaselist e {e_{\snil}} {x} {xs} {e_{\sconskw}}}} = \\
\cpot*{\ccaselist
  {\cpot{\ctrans e}}
  {\ctrans{e_{\snil}}}
  {x}{xs}
  {\ctrans{e_{\sconskw}}}
}.
\end{multline*}

\section{Models}
\label{sec:models}

It is the interpretation of syntactic recurrences that correspond to
informally-extracted recurrences and provide actual bounds
on the cost of source language programs, so let us define the notion of
model for the recurrence language and see how models arise in cost analysis.
For this discussion, syntax refers to the recurrence language unless
explicitly stated otherwise.

A \emph{type frame} is a set~$\model A$ of preordered sets 
along with a denotation function
$\den\cdot : \synType \to \model A$.
For a type frame $\model A$ and a type context~$\gamma$, a
\emph{$\gamma$-environment} is a function 
$\eta:\synVar\to\bigunion \model A$ such that for all $x\in\dom\gamma$,
$\eta(x)\in\den{(\gamma(x))}$.  We write $\synEnv_\gamma$ for the set of
$\gamma$-environments and $\synTerm_\gamma$ for the set of type derivations
with type context~$\gamma$.
A type frame is an \emph{interpretation} if there is a denotation function
$\den\cdot : \synTerm_\gamma \to \synEnv_\gamma \to \bigunion\model A$
such that if $\typejudge\gamma e \sigma$ and $\eta\in\synEnv_\gamma$, then
$\den e\eta\in\den\sigma$.  An interpretation is a \emph{model} if it
satisfies the size order rules of Figure~\ref{fig:size-order}.  That is,
writing the preorder on $\den\sigma$ as $\szleq_{\den\sigma}$, if
$\szjudge\gamma{e}{e'}{\sigma}$ and $\eta\in\synEnv_\gamma$, then
$\den[\eta]{e} \szleq_{\den\sigma} \den[\eta]{e'}$.

\begin{wrapfigure}{R}{.45\linewidth}
\begin{callout}
Is this an appropriate use of the term ``adequacy?'' It is saying something
like denotational behavior implies operational behavior.  But it isn't
saying that denotational equivalence implies syntactic equivalence (of some
kind).  Note that \citet{kavvos-et-al:popl2020} prove this sort of result
via actual adequacy for the recurrence language.
\end{callout}
\end{wrapfigure}

Our interest in models stems from a kind of adequacy-like result for costs:
we want to show that if $\den{\ccost{\ctrans e}}\leq n$ in some model, then
$\eval e v c$ for some $c\leq n$---i.e., if the semantic recurrence yields a
finite cost, then the program evaluates with no greater cost.  We might be
able to prove this unconditionally, but that seems to be complicated by the
fact that evaluation cost is determined by placement of the $\stick{}$
operator.  For now let us consider a relatively simple proof that relies on
doing so in a ``reasonable'' way, which in turn lets us take into account
the actual size of the evaluation derivation.

\begin{defi}
A derivation~$\Pi$ of $\ieval e c$ is \emph{size-maximal} if it is the
maximum size of any derivation of~$\ieval e {c'}$ (for any~$c'$).
\end{defi}

\begin{lem}
Suppose $\Pi$ is a size-maximal derivation of~$\ieval e c$.  Then
there is a derivation of~$\eval e v c$ for some~$v$ that is no smaller
than~$\Pi$
\end{lem}
\begin{proof}
The proof is by induction on~$\Pi$, breaking into cases according to the
last rule.

Suppose $\Pi$ is $\ndAXC{$\ievalold e 0 0$}\DisplayProof$.  If $e$ is a value,
then the claim follows from Prop.~\ref{prop:value-eval}.  If $e$ is not a
value, then there is another derivation of~$\ieval e c$ using one of the
other rules, contradicting size-maximality of~$\Pi$.

The proofs for the other rules all follow the same pattern; we do the
pairing rules as an example.
\begin{proofcases}
\item[$
    \AXC{$\Pi_0$}
    \noLine
  \ndUIC{$\ieval{e_0}{c_0}$}
\ndUIC{$\ieval{\spair{e_0}{e_1}}{c_0}$}
\DisplayProof
$]
Size-maximality of~$\Pi$ implies the same of the hypothesis, so by the
induction hypothesis there is a derivation~$\Pi'$
of~$\eval {e_0} {v_0}{c_0}$ that
is no smaller than~$\Pi_0$.
But that means that there is a larger derivation
of~$\ieval{\spair{e_0}{e_1}}{c_0}$ of the form
\begin{prooftree}
    \AXC{$\Pi'$}
    \noLine
  \ndUIC{$\eval{e_0}{v_0}{c_0}$}
  \ndAXC{$\ieval{e_1}{0}$}
\ndBIC{$\ieval{\spair{e_0}{e_1}}{c_0}$}
\end{prooftree}
contradicting size-maximality of~$\Pi$.

\item[$
  \AXC{$\eval{e_0}{v_0}{c_0}$}
    \AXC{$\Pi_1$}
    \noLine
  \ndUIC{$\ieval{e_1}{c_1}$}
\ndBIC{$\ieval{\spair{e_0}{e_1}}{c_0+c_1}$}
\DisplayProof
$]
As before, size-maximality of~$\Pi$ implies that by the IH there is a
derivation~$\Pi'$ of~$\eval {e_1} {v_1} {c_1} $ that is no smaller
than~$\Pi_1$, and hence
\begin{prooftree}
  \AXC{$\eval{e_0}{v_0}{c_0}$}
    \AXC{$\Pi'$}
    \noLine
  \ndUIC{$\eval{e_1}{v_1}{c_1}$}
\ndBIC{$\eval{\spair{e_0}{e_1}}{\spair{v_0}{v_1}}{c_0+c_1}$}
\end{prooftree}
is the desired derivation.
\end{proofcases}
\end{proof}

\begin{prop}
\label{prop:eval-or-diverge}
For all closed~$e$, either (1)~there are~$v$ and $c$ such that
$\eval e v c$, or (2)~for all~$k$ there is some~$c$ and a derivation 
of~$\ieval e {c} $ of size at least~$k$.
\end{prop}

\begin{wrapfigure}{r}{3.5in}
\begin{callout}
This is sort of a weird definition.  Values (including function definitions)
are not sensibly ticked, because if $v$ is a value, then $\eval v v 0$.
There is probably a more logical-relation-y kind of definition that does the
right thing.  Regardless, of course, being well-ticked will not be
decidable.
\end{callout}
\end{wrapfigure}

Our use of the ticking mechanism to count cost means that we cannot conclude
anything about the cost of the incomplete evaluations when~$e$ does not
evaluate to a value, since~$e$ could be a non-terminating computation with
no ticked subexpressions at all.  Of course, that is not a particularly
well-ticked program.  

\begin{defi}
Define~$e$ to be \emph{sensibly ticked} if there are
constants~$a>0$ and~$b>0$ such that whenever there is a size~$k$ 
derivation of~$\ieval e c$, $ak+b \leq c$.
\end{defi}

\begin{defi}
For a model~$\model A$, we write $n$ 
for~$\den{\overbrace{1+\dots+1}^{\text{$n$ times}}}$.
A model~$\model A$ is \emph{cost-standard} if $\szleq_\C$ is a partial
order and $0 < 1 < 2 < \dotsb$.
\end{defi}

\begin{prop}[Adequacy for costs]
\label{prop:semantic-adequacy}
If $e$ is closed and
sensibly ticked and $\model A$ is a cost-standard model and there is some~$n$
such that $\den{\ccost{\ctrans e}} \leq n$,
then there are~$c\leq n$ and~$v$ such that $\eval e v c$.
\end{prop}
\begin{proof}
If there is no complete evaluation of~$e$, then 
by Prop.~\ref{prop:eval-or-diverge}, for every~$k$ there is~$c_k$ such that
there is a derivation of~$\ieval e {c_k}$ of size $\geq k$.
By sensible ticking there are fixed $a>0$ and~$b>0$ such that $ak+b\leq c_k$
for all~$k$, and hence there are arbitrarily large~$c$
such that $\ieval e c$.  So by the Soundness Theorem, there
are arbitrarily large~$c$
such that $c\leq \ctrans{e}_c \leq n$, which contradicts the assumption
that~$\model A$ is cost-standard.
So by Prop.~\ref{prop:eval-or-diverge} again,
there is a derivation of~$\eval e v c$ for some~$c$, and hence by
the Soundness Theorem, $c\leq \ctrans{e}_c \leq  n$.
\end{proof}

\begin{wrapfigure}{r}{.45\linewidth}
\begin{callout}
Is there anything interesting to talk about for a class of models with the
same interpretation of~$\C$?
\end{callout}
\end{wrapfigure}

So establishing a cost bound on a source language program~$e$ consists of
defining a model~$\model A$ and showing that the semantic recurrence 
$\den{\ctrans e}$ extracted from~$e$ is bounded in~$\model A$.
When a collection of models all agrees on $\den\C$, the models differ in how
they yield an interpretation of size/potential.  Analyzing a program, then,
typically involves understanding the appropriate notion of size.  As we will
see, understanding ``appropriate notion of size'' is necessary not only for
values of inductive type, but also for other types such as sums and
products.

On its own, this adequacy result is probably less useful than it appears,
because in an arbitrary model it is probably difficult to prove that there is~$n$ such that
$\den{\ccost{\ctrans e}} \leq n$.  
The issue is that~$\ctrans e_c$ is a complex expression,
and so computing $\den{\ccost{\ctrans e}}$ probably
relies on reducing subexpressions along the lines of
$\ctrans e_c = E_0 \to E_1 \to E_2\to\dotsb\to E$ and computing~$\den E$.
But the size order tells us
that $\den{E_i} \szgeq \den{E_{i+1}}$, so even if~$\den E = m$, 
that only tells us that
$\den{\ccost{\ctrans e}} \geq m$, 
not $\den{\ccost{\ctrans e}} \leq n$ for some~$n$.  But the models we
define in the sequel satisfy most of the $\beta$ axioms in a very strong
way:  each such axiom that asserts that $e \szleq e'$ is valid because $e$
and $e'$ are in fact equal.  We use that to simplify the semantic
recurrences.  As long as our simplifications only use those $\beta$ axioms,
we know that the simplified recurrences are equal to the originally
extracted recurrences, and thus Proposition~\ref{prop:semantic-adequacy}
gives us useful information.

\section{A constructor-counting model for merge sort}

Our first goal is to analyze the standard merge sort algorithm on lists,
counting item comparisons.  The usual analysis is in terms of the length of
the argument list, and hence we need a model in which the denotation of
a value of type $\clist\sigma$ is its length.  To do so, we define a model
that counts the number of ``main non-nullary constructors'' in each
inductive value.
That is, we
interpret a recursive value~$v$ of type~$\delta$
as the number of non-nullary $\cfoldkw$ constructors in~$v$ used to
construct type~$\delta$ subvalues.  Thus the potential (size) of a source
language list would be the
number of $\sconskw$ constructors, the potential of a source language
tree would be the number of $\snodekw$
constructors, etc.
To simplify the description, we restrict ourselves to the
setting in which recursive types in the source language
have the form~$\slfp\alpha F$, where~$F$
is given by the grammar
\[
F ::= \alpha \mid \sigma \mid \ssum{F_0}{F_1} \mid \sprod{F_0}{F_1}
\]
and the $\sigma$ production is restricted to closed~$\sigma$.  We restrict
recursive types in the recurrence language similarly and observe that the
translation functions in Figure~\ref{fig:rec-extraction} map the restricted
source language to the restricted recurrence language.  As mentioned
earlier, in this restricted setting, we require $\sfunkw$ and $\cfixkw$ for
recursive function definitions.

\subsection{The constructor counting model}

\begin{figure}
\begin{align*}
\den\alpha\eta &= \eta(\alpha) \\
\den\C\eta 
  &= \N^\infty & x\szleq y & \Iff x\szleq_{\N} y \vee y = \infty \\
\den\cunit\eta &= \set{*} & x\leq y & \Iff x = y \\
\den{\csum{\sigma_0}{\sigma_1}}\eta
  &= \powerset(\den{\sigma_0}\eta\disjunion\den{\sigma_1}\eta)
  &  X \szleq Y &\Iff \forall x\in X\exists y\in Y. \\
  &  &          & \qquad x\szleq_{\den{\sigma_0}\eta\disjunion\den{\sigma_1}\eta} y \\
\den{\cprod{\sigma_0}{\sigma_1}}\eta
  &= \den{\sigma_0}\eta\cross\den{\sigma_1}\eta
  &  (x_0, x_1) \szleq (x_0', x_1') &\Iff 
     x_i\szleq_{\den{\sigma_i}\eta} x_i', \\
  &  & & \qquad i=0,1 \\
\den{\carr{\sigma_0}{\sigma_1}}\eta
  &= \setst{f : \den{\sigma_0}\eta \to \den{\sigma_1}\eta}
           {\text{$f$ is $\szleq$-monotone}}
  &  f \leq g &\Iff \forall x.f(x)\szleq g(x) \\
\den{\clfp\alpha\sigma}
  &= \N^\infty
  &  x\szleq y & \Iff x\szleq_{\N} y \vee y = \infty \\
\den{\ccpy\sigma}\eta
  &= \C\cross\den\sigma\eta
  &  (c, a) \szleq (c', a') &\Iff c \szleq_{\C} c' \And a\szleq_\sigma a'
\end{align*}
\caption{The interpretation of types in the constructor size model.}
\label{fig:csize-model-types}
\end{figure}

The interpretation of types is given in Figure~\ref{fig:csize-model-types},
with the size order shown as well (we will discuss the information order
needed to interpret recursive function definitions momentarily).
We assume that $\eta$ is a map from type variables to
semantic types.  $\N^\infty = \N\union\set{\infty}$, where $\N$ is the
natural numbers with the usual order,
and $\powerset(\cdot)$ is the powerset operator.  
We will frequently write~$\den\sigma$ instead of $\den\sigma\eta$
when $\eta$ is irrelevant.
A key feature of this type interpretation is that $\den\sigma$ is always
a complete lattice under $\szleq_{\den\sigma}$, 
a fact that is straightforward to verify.
We interpret products and arrows in the usual way, writing $\smfst$
and~$\smsnd$ for left- and right-projection, but sums are a bit more
complex.  For ordered sets~$X_0$ and~$X_1$, define the disjoint union
$X_0\disjunion X_1$ as usual, with injection functions~$\sminj i:X_i\to
X_0\disjunion X_1$, and define $\szleq_{X_0\disjunion X_1}$ by
$\sminj i x\szleq_{X_0\disjunion X_1} \sminj i x' \Iff x\szleq_{X_i} x'$.
Of course, $X_0\disjunion X_1$ is not a complete lattice under this order.
One solution is to add a top element, but this ends up resulting in very
weak solutions to extracted recurrences; see
\citet{danner-licata:jfp22} for a discussion.  This same issue arises
in static program analysis, and the solution is the same:  use (a subset of)
$\powerset(X_0\disjunion X_1)$ instead.
Previously we have used
order ideals (i.e., downward-closed sets) with the inclusion
ordering.  Here we will use all subsets, ordered by
$X\szleq Y$ if every element of~$X$ is bounded above by some element of~$Y$.
The reason for this more complex ordering is that for arbitrary subsets,
desired monotonicity properties are lost in the inclusion order.  For
example, the implication
$a\leq a' \Implies \set{\sminj 0 a}\subseteq\set{\sminj 0 a'}$ fails, where
$\sminj 0$ is injection into the left summand.  We will have occasion to
interpret products similarly, but because calculating the semantic
recurrences is messier with these interpretations, we avoid them unless they
are needed.

As already mentioned, for every~$\sigma$,
$\den\sigma$ is a complete upper semilattice with respect to~$\szleq$
(i.e., the least upper bound $\bigszlub X$ exists for every
$X\subseteq\den\sigma$).
It is a standard fact
that a complete upper semilattice is a complete lattice, with
$\bigszglb X = \bigszlub\setst{a}{\forall x\in X.a\leq x}$.
We observe that 
$\bigszlub_{\den{\csum{\sigma_0}{\sigma_1}}} \mathcal X = 
\bigunion \mathcal X$, but in general
$\bigszglb_{\den{\csum{\sigma_0}{\sigma_1}}}\mathcal X\not=
\bigintersection\mathcal X$ (cf.\ the complete
upper semilattice of open sets of a topological space).
In order to interpret~$\cfixkw$, we also need an information order~$\infleq$
on each type such that $(\den\sigma,\infleq_{\den\sigma})$ is
chain-complete.
We set $\mathord\infleq = \mathord\szgeq$ for all types, so every type is
also a complete lattice with respect to~$\infleq$, and hence trivially
chain-complete.  However, we only require that arrow types be interpreted by
monotone functions, not $\infleq$-continuous functions.\footnote{That is
because so far, I've been unable to prove that the term denotation function
is $\szgeq$-continuous.}  Continuity is not really necessary, because in a
complete lattice, monotone functions still have a least fixpoint.
More precisely,
if $(D,\infleq)$ is a complete lattice (which necessarily has a bottom
element~$\bot_\infleq$)
and $f:D\to D$ a $\infleq$-monotone
function, define~$x_\alpha$ for ordinals~$\alpha$ by
\[
x_0 = \bot
\qquad
x_{\alpha+1} = f(x_\alpha)
\qquad
x_\lambda = \bigszlub\nolimits_{\infleq}\setst{x_\gamma}{\gamma<\lambda}
\]
Then there is some (in fact, exactly one) $\alpha$
such that $x_{\alpha+1} = x_\alpha$; $x_\alpha$ is the least fixpoint
of~$f$ with respect to~$\infleq$, 
and that will be the value of $\smfix_D(f)$.  
In practice we will see that for the functions~$f$ that we
consider, $x_\omega$ is a fixpoint, and hence the least fixpoint.
In our setting with $\infleq = \szgeq$, $\bot_{\infleq}$ is the top element
with respect to~$\szleq$---that is, the least defined semantic value is the
semantic value that provides the least useful size bound.  Similarly,
$\bigszlub_{\infleq}$ is $\bigszglb_{\szleq}$, and in particular an
$\infleq$-increasing chain is a $\szleq$-decreasing chain---that is is, a
sequence of more defined semantic values is a sequence of better bounds.

\begin{figure}
\begin{align*}
\den{x}\eta &= \eta(x) \\
\den{\ctriv}\eta &= * \\
\den{0}\eta &= 0 \\
\den{1}\eta &= 1 \\
\den{e_0+e_1}\eta &= \den{e_0}\eta + \den{e_1}\eta \\
\den{\cinj i e}\eta &= \set{\sminj i(\den{e}{\eta})} \\
\den{\ccase e x {e_0} x {e_1}}\eta
  &= \bigszlub\Bigl(
                \setst{\den{e_0}{\envext\eta x a}}{\sminj 0(a)\in\den{e}\eta} 
                 \union \\
  &\qquad\qquad  \setst{\den{e_1}{\envext\eta x a}}{\sminj 1(a)\in\den{e}\eta}
               \Bigl) \\
\den{\cpair{e_0}{e_1}}{\eta}
  &= (\den{e_0}\eta, {\den{e_1}\eta}) \\
\den{\clet{\cpair{x_0}{x_1}}{e'}{e}}{\eta}
  &= \den{e}{\envext\eta{x_0,x_1}{\smfst(\den e \eta),\smsnd(\den e\eta)}} \\
\den{\clam x e}{\eta}
  &= \llambda a.\den{e}{\envext\eta x a} \\
\den{\capp{e_0}{e_1}}{\eta}
  &= \left(\den{e_0}{\eta}\right)\left(\den{e_1}{\eta}\right) \\
\den{\cfix x e}{\eta}
  &= \smfix_{\szgeq}(\llambda a.\den e {\envext\eta x a}) \\
\den{\cfold[\clfp\alpha F] e}{\eta} 
  &= \smsucc(\smcsize_{F}(\den e\eta) \\
\den{\cunfold[\clfp\alpha F] e}{\eta} 
  &= \bigszlub\setst{a}{\smsucc(\smcsize_F(a)) \szleq \den e \eta}
\\
\den{\cval e}\eta &= (0, \den e\eta) \\
\den{\cbind x {e'} e}{\eta}
  &= (\smfst(\den e {\envext\eta x {p'}}) + \den{e'}{\eta}, 
      \smsnd(\den e {\envext\eta x {p'}})) \\
\den{\cincr e}\eta
  &= (\smfst(\den e\eta) + 1, \smsnd(\den e\eta)) \\
\den{\ccost e}\eta
  &= \smfst(\den e \eta) \\
\den{\cpot e}\eta
  &= \smsnd(\den e \eta)
\end{align*}
\caption{The interpretation of terms in the constructor size model.  The
semantic functions~$\smsucc$ and~$\smcsize_F$ are defined in
Figure~\ref{fig:csize-model-csize-fn}.}
\label{fig:csize-model-terms}
\end{figure}

\begin{figure}
\begin{align*}
\smcsize_F &: F[\N^\infty] \to \set{\bot}\union \N^\infty \\
\smcsize_\alpha(n) &= n \\
\smcsize_\sigma(a) &= \bot \\
\smcsize_{\csum{F_0}{F_1}}(X_0\disjunion X_1) 
  &= \bigszlub\Bigl(
       \setst{\smcsize_{F_0}(a)}{a\in X_0}
       \union
       \setst{\smcsize_{F_1}(a)}{a\in X_1}
     \Bigr) \\
\smcsize_{\cprod{F_0}{F_1}}(a_0, a_1)
  &= \smcsize_{F_0}(a_0) + \smcsize_{F_1}(a_1)
\\ \\
\smsucc &: \set{\bot}\union\N^\infty\to\N^\infty \\
\smsucc(\bot) &= 0 \\
\smsucc(n) &= n+1 \\
\smsucc(\infty) &= \infty
\end{align*}
\caption{The semantic constructor size function $\smcsize$.}
\label{fig:csize-model-csize-fn}
\end{figure}

The interpretation of terms is given in Figure~\ref{fig:csize-model-terms}.
It is straightforward to prove that for all~$e$, $\eta$, and~$x$,
$\den e {\envext\eta x \cdot}$ is monotone with respect to~$\leq$ (and hence
$\szgeq$), and so the preceding discussion justifies the definition
of~$\den{\cfix x e}\eta$.  What is left is the interpretations 
of $\cfoldkw$ and $\cunfoldkw$.
Remember that our goal is to interpret a
value of type~$\delta = \slfp\alpha F$ as the
number of $\cfoldkw$ applications in the value whose arguments do not
contain any subvalues of type~$\delta$.  
To do so we define an auxiliary
function~$\smcsize_F$ that maps $F[\N^\infty]\to \set{\bot}\union\N^\infty$,
where $\bot\leq y$ for all~$y\in\N^\infty$ and $x+\bot = \bot+x = x$ for
all~$x\in\Ninfty$.  The idea is that to
compute~$\den{\cfold[\delta] e}{}$, we inductively compute the size 
of~$\den e$ and add~$1$.  But we want to ensure that if~$e$ has no type-$\delta$
subterms, then the result has total size~$0$.  Thus we ``shift the count
down by one'' (that is the role of~$\bot$) and ``shift back up'' after
counting (that is the role of~$\smsucc$).  The definition can be adapted to
yield height (e.g., for trees) by replacing the sum in the
$\cprod{F_0}{F_1}$ case with maximum.

Since we interpret product by ordinary cartesian product and $\ccpy\cdot$ as
$\cprod{\C}{\cdot}$, the following fact is trivial to verify this in this
model:

\begin{prop}
\label{prop:cpxy-equals-cost-pot-pair}
For all $z\in\den{\ccpy\sigma}$, $z = (\smfst\,z, \smsnd\,z)$.  In
particular, if $E\oftype\ccpy\sigma$, then
$\den E\eta = (\den{\ccost E}\eta, \den{\cpot E}\eta)$.
\end{prop}

\begin{figure}
\begin{align*}
\withcost{e}{c} &= \cpair{c}{e} \\
\plusc{c}{e} &= \cpair{c+\ccost e}{\cpot e}
\end{align*}
\caption{Notation for describing costs in the constructor-counting model.}
\label{fig:extra-cost-notation}
\end{figure}

Since a complexity in this model is a pair consisting of a cost and a
potential, we introduce the notation in
Figure~\ref{fig:extra-cost-notation}; though wordier than writing pairs, we
find it easier to read.  We will use this notation extensively in the
recurrence language syntax, and for the remainder of this section, when we
write syntactic expressions, we are really referring to their denotations in
the constructor counting model.

Let us make some observations about this model:

\begin{enumerate}

\item The axioms \infruleref{beta-plus}, \infruleref{beta-times},
\infruleref{beta-to}, and \infruleref{beta-fix} are all satisfied because
in each case, if the axiom asserts~$e\szleq e'$, then in fact
$\den e = \den e'$.

\item The intent behind this model is that a list is interpreted as the
number of $\sconskw$ constructors in it.  To verify this, we first compute
the denotations of $\cnil$ and $\cconskw$ expressions:
\begin{align*}
\den{{\cnil}}
  &= \den{{\cfold{(\cinj 0 \ctriv)}}} \\
  &= \smsucc(\smcsize_{\csum\cunit{\cprod\sigma\alpha}}(
       \den{\cinj 0 \ctriv}
     ) \\
  &= \smsucc(\smcsize_{\csum\cunit{\cprod\sigma\alpha}}(
       \set{\sminj 0 *}
     ) \\
  &= \smsucc(\bot) \\
  &= 0
\end{align*}
and
\begin{align*}
\den{{\ccons{e}{es}}}
  &= \den{
         {\cfold{(\cinj 1 {\cpair{{\den e}}{{\den{es}}}})}}
     } \\
  &= \smsucc(\smcsize_{\csum\cunit{\cprod\sigma\alpha}}(
       \den{\cinj 1 {\cpair{{\den e}}{{\den{es}}}}}
     ) \\
  &= \smsucc(\smcsize_{\csum\cunit{\cprod\sigma\alpha}}(
       \set{\sminj 1 {(\den{e}, \den{es})}}
     ) \\
  &= \smsucc(\smcsize_{{\cprod\sigma\alpha}}(
       (\den{e}, \den{es})
     ) \\
  &= \smsucc(
       \smcsize_\sigma(\den{e}) +
       \smcsize_\alpha(\den{es})
     ) \\
  &= \smsucc(
       \bot +
       \den{es}
     ) \\
  &= \smsucc(
       \den{es}
     ) \\
  &= 1 + \den{es} & &  \qquad \text{(because $\den{e'}\not=\bot$ for any~$e'$)}
\end{align*}

We can now use this to simplify the extraction of $\snil$ and $\sconskw$
expressions.  We use
Proposition~\ref{prop:cpxy-equals-cost-pot-pair} to simplify the
denotation of many extracted recurrences by writing them as cost-potential
pairs and using the calculations in Section~\ref{sec:lists}.  Thus:
\begin{align*}
\den{\ctrans{\snil}}
  &= \den{\cval\cnil} \\
  &= \withcost{\den{\cnil}}{0} \\
  &= \withcost{0}{0}
\end{align*}
and
\begin{align*}
\den{\ctrans{\scons{e}{es}}}
  &= \withcost
       {\den{\cpot{\ctrans{\scons{e}{es}}}}}
       {\den{\ccost{\ctrans{\scons{e}{es}}}}} \\
  &= \withcost
       {\den{\ccons{\cpot{\ctrans e}}{\cpot{\ctrans{es}}}}}
       {\den{\ccost{\ctrans e} + \ccost{\ctrans{es}}}} \\
  &= \withcost
       {1 + \den{\cpot{\ctrans{es}}}}
       {\den{\ccost{\ctrans e} + \ccost{\ctrans{es}}}}.
\end{align*}

\item We next compute the values of $\cunfoldkw$ expressions, where we will
be a little informal and write elements of~$\Ninfty$ as though they are
syntax:
\[
\den{\cunfold 0} = \set{*}\disjunion\emptyset
\qquad
\den{\cunfold n} = \set{*}\disjunion\setst{j}{j < n}
\qquad
\den{\cunfold \infty} = \set{*}\disjunion\Ninfty
\]

\item 
Now we can compute the denotations of $\ccasekw_{\clist\sigma}$ expressions.
\begin{multline*}
\den{\ccaselist 0 {e_\cnil} {x} {xs} {e_\cconskw}}\eta \\
  \begin{aligned}[t]
  &= \den{
       \ccase 
         {\cunfold 0} 
         \cwild 
         {e_\cnil} 
         z
         {\clet {\cpair x {xs}} z {e_\cconskw}}
     }\eta \\
  &= \bigszlub\Bigl(
       \setst{\den{e_{\cnil}}\eta}{\cwild\in\set{*}}
       \union{} \\
  &\qquad\qquad\setst
         {\den{\clet{\cpair x {xs}} z {e_{\cconskw}}}
              {\envext\eta z {(a, j)}}}
         {a\in\den\sigma, j < 0}
     \Bigr) \\
  &= \den{e_{\cnil}}\eta
  \end{aligned}
\end{multline*}
and for $n>0$,
\begin{multline*}
\den{\ccaselist n {e_\cnil} {x} {xs} {e_\cconskw}}\eta \\
  \begin{aligned}[t]
  &= \den{
       \ccase 
         {\cunfold n} 
         \cwild 
         {e_\cnil} 
         z
         {\clet {\cpair x {xs}} z {e_\cconskw}}
     }\eta \\
  &= \bigszlub\Bigl(
       \setst{\den{e_{\cnil}}\eta}{\cwild\in\set{*}}
       \union{} \\
  &\qquad
       \setst[\left]
         {\den{\clet{\cpair x {xs}} z {e_{\cconskw}}}
              {\envext\eta z {(a, j)}}}
         {
           a\in\den\sigma, j < n
         }
     \Bigr) \\
  &= \bigszlub\Bigl(
       \set{\den{e_{\cnil}}\eta}
       \union
       \setst
         {\den{\clet{\cpair x {xs}} z {e_{\cconskw}}}
              {\envext\eta z {(a, j)}}}
         {
           a\in\den\sigma, j < n
         }
     \Bigr) \\
  &= \bigszlub\Bigl(
       \set{\den{e_{\cnil}}\eta}
       \union
       \set
         {
           \den
             {\clet{\cpair x {xs}} z {e_{\cconskw}}}
             {\envext\eta z {(\infty,n-1)}}
         }
     \Bigr) \\
  & \qquad\text{(by monotonicity of $\den\cdot$)} \\
  &= \bigszlub\Bigl(
       \set{\den{e_{\cnil}}\eta}
       \union
       \set
         {
             \den{e_{\cconskw}}{\envext\eta{x,xs}{\infty,n-1}}
         }
     \Bigr) \\
  &= \den{e_{\cnil}}\eta \bmax 
     \den{e_{\cconskw}}{\envext\eta{x,xs}{\infty,n-1}}
  \end{aligned}
\end{multline*}

\noindent
Similarly,
\[
\den{\ccaselist \infty {e_\cnil} {x} {xs} {e_\cconskw}}\eta =
\den[\eta]{e_{\cnil}}\bmax
\den[\envext\eta{x,xs}{\infty,\infty}]{e_{\cconskw}}.
\]

\item 
We also observe that~\infruleref{mon} holds for arbitrary contexts in this
model.

\begin{prop}
\label{prop:mon-for-all-contexts}
The following inference rule is valid in this model for any term
context~$\elimctx{}$:
\[
  \AXC{$e\szleq e'$}
\ndUIC{$\elimctx{e}\szleq \elimctx{e'}$}
\DisplayProof
\]
\end{prop}

\end{enumerate}

Thanks to these observations, we shall use the equations in
Figure~\ref{fig:cc-model-eqns} when reasoning about complexity language
expressions to be interpreted in this model.

\begin{figure}
\begin{align*}
\subst{e_i}{e}{x}
  &= \ccase{\cinj i e} x {e_0} x {e_1} \\
\subst e {e_0,e_1} {x_0,x_1}
  &= \clet {\cpair{x_0}{x_1}} {\cpair{e_0}{e_1}} e \\
\subst e {e_1} x 
  &= \capp{(\clam x e)}{e_1} \\
\subst e {\cfix x e} {x} 
  &= \cfix x e \\
\\
\cbind{p}{e'}{e}
  &= \plusc{\ccost{e'}}{\subst{e}{\cpot{e'}}{p}} \\
\cbind{p}{\withcost{e'}{c}}{e}
  &= \plusc{c}{\subst{e}{e'}{p}} \\
\cbind{p_0,\dots,p_{n-1}}{e_0,\dots,e_{n-1}}{e}
  &= \subst{e}{e_0',\dots,e_{n-1}'}{p_0,\dots,p_{n-1}} \\
  &  \qquad\qquad \text{($e_i = \withcost{e_i'}{0}$ or $e_i = \cval{e_i'}$)}
     \\
\plusc{0}{e} 
  &= e \\
\\
\ctrans{\scons{e}{es}}
  &= \withcost
       {1+\cpot{\ctrans{es}}}
       {\ccost{\ctrans{e}}+\ccost{\ctrans{es}}} \\
\ctrans{\scaselist{e}{e_{\snil}}{y}{ys}{e_{\sconskw}}}
  &= \plusc*
       {\ccost{\ctrans e}}
       {(
         \ccaselist
           {\cpot{\ctrans e}}
           {\ctrans{e_{\snil}}}
           {y}{ys}
           {\ctrans{e_\sconskw}}
       )}
\end{align*}
\caption{Equations valid in the constructor-counting model.}
\label{fig:cc-model-eqns}
\end{figure}

\subsection{The analysis of merge sort}

\begin{figure}
\[
\begin{array}{l}
\sfun*{split}
       {xs}
       {\scaselist*{xs}
               {\spair\snil\snil}
               {y}{ys}
               {\scaselist*{ys}
                       {\spair{[y]}{\snil}}
                       {z}{zs}
                       {\slet*{\spair{as}{bs}}
                              {\sapp{split}{zs}}
                              {\spair{\scons y {as}}{\scons z {bs}}}}}} \\
\sfun*{merge}{xsys}
      {\slet*{\spair{xs}{ys}}{xsys}
             {\scaselist*{xs}
                     {ys}
                     {x'}{xs'}
                     {\scaselist*{ys}
                             {xs}
                             {y'}{ys'}
                             {\sif*{\stick*{x'\leq y'}}
                                  {\scons{x'}{\sapp{merge}{\spair{xs'}{ys}}}}
                                  {\scons{y'}{\sapp{merge}{\spair{xs}{ys'}}}}}}}}
                                  \\
\sfun*{msort} 
     {xs}
     {\scaselist*{xs}
            {\snil}
            {y}{ys}{\scaselist*{ys}
                          {[y]}
                          {\swild}{\swild}
                          {\slet*{\spair{ws}{zs}}{\sapp{split}{xs}}
                                {\sapp{merge}
                                      {\spair{\sapp{msort}{ws}}
                                             {\sapp{msort}{zs}}}}}}}
\end{array}
\]
\caption{Merge sort}
\label{fig:merge-sort}
\end{figure}

\begin{figure}
{\small
\begin{align*}
& \cpot{\ctrans{\sfun{split}{xs}{\cdots}}} = \\
& \cfix*{split}{
    \clam{xs}{
      \ccaselist*
      {xs}
      {
        \czcost{\cpair 0 0}
      }
      {y}{ys}
      {
        \ccaselist*
          {ys}
          {
            \czcost{\cpair{1}{0}}
          }
          {z}{zs}
          {
            \plusc*
              {\ccost*{\capp{split}{zs}}}
              {
                \clet*{\cpair{as}{bs}}{\cpot*{\capp{split}{zs}}}{
                  \czcost{\cpair{1+as}{1+bs}}
                }
              }
          }
      }
    }
  } \\
\\
& \cpot{\ctrans{\sfun{merge}{xsys}{\cdots}}} \\
& \cfix*{merge}{
    \clam{xsys}{
      \clet*{\cpair{xs}{ys}}{xsys}{
        \ccaselist*
          {xs}
          {\czcost{ys}}
          {x'}{xs'}
          {
            \ccaselist*
              {ys}
              {\czcost{xs}}
              {y'}{ys'}
              {
                \plusc*
                  1
                  {
                    \cif*
                      {x'\leq y'}
                      {
                        \clet*{\chi}{\capp{merge}{\cpair{xs'}{ys}}}{
                          \withcost
                            {1+\cpot\chi}
                            {\ccost\chi}
                        }
                      }
                      {
                        \clet*{\chi}{\capp{merge}{\cpair{xs}{ys'}}}{
                          \withcost
                            {1+\cpot\chi}
                            {\ccost\chi}
                        }
                      }
                  }
              }
          }
      }
    }
  } \\
& \cpot{\ctrans{\sfun{msort}{xs}{\cdots}}} = \\
& \cfix*{msort}{
    \clam{xs}{
      \ccaselist^
        {xs}
        {\czcost 0}
        {y}{ys}
        {
          \ccaselist^
            {ys}
            {\czcost{1}}
            {\cwild}{\cwild}
            {
              \plusc*
                {\ccost*{\capp{split}{xs}}}
                {
                  \clet*{\cpair{ws}{zs}}{\cpot*{\capp{split}{xs}}}{
                    \plusc*
                    {
                      \ccost*{\capp{msort}{ws}} +
                      \ccost*{\capp{msort}{zs}}
                    }
                    {
                      \capp
                        {merge}
                        {
                          \spair
                            {\cpot*{\capp{msort}{ws}}}
                            {\cpot*{\capp{msort}{zs}}}
                        }
                    }
                  }
                }
            }
        }
    }
  }
\end{align*}
}
\caption{Syntactic recurrences for merge sort.}
\label{fig:merge-sort-syn-recs}
\end{figure}

Merge sort over $\slist\sint$ is defined in Figure~\ref{fig:merge-sort}.  We
assume~$\sint$ is defined as a large enumerated type with~$\leq$ defined by
table lookup, and we only count $\sint$ comparisons.  The syntactic
recurrences are given in Figure~\ref{fig:merge-sort-syn-recs}.  These are
recurrences that have been simplified making use of equations valid in the
model.  The derivations are given in
Appendix~\ref{app:merge-sort-syn-rec-derivations}.  To reduce clutter, these
simplifications make use of equations valid in the constructor counting
model---that is, the reader should pretend as though there are denotation
brackets around every term in that figure.

Let us now write $split$ for $\den{\cfix{split}{\cdots}}$,
$split_c = \smfst\comp split$ and $split_p = \smsnd\comp split$.
It is the last two functions we care most about.
Since our interest is in showing that our method results in the same
recurrences as those derived informally, we shall use the final expression in 
Figure~\ref{fig:split-syn-rec} to write down the recurrence relations
satisfied by $split_c$ and~$split_p$.  To that end, let $F$ be the
functional defined by that expression so that $split = \smfix(F)$.  
Then, again imagining denotation
brackets wherever appropriate, we see that
\begin{multline*}
split_c 
  = \smfst\comp split
  = \smfst\comp\cpot{\ctrans{\sfun{split}{xs}{\cdots}}}
  = \smfst\comp \smfix(F)
  = \smfst\comp F(\smfix(F)) \\
  = \smfst\comp(F\,split)
  = \smfst\comp \den{\lambda xs.\cdots}{\env{\envbind{split}{split}}}.
\end{multline*}
So when interpreting the term defining~$F$,
\[
\den{\ccost*{\capp{split}{zs}}}
  = (\smfst\comp split)\den{zs}
  = split_c(\den{zs})
\]
and similarly $\den{\cpot*{\capp{split}{zs}}} = split_p(\den{zs})$.
This lets us conclude that
\begin{align*}
split_c(0) &= 0 \\
split_c(1) &= 0 \bmax 0 = 0 \\
split_c(n) &= 0 \bmax 0 \bmax split_c(n-2) + 0 = 0 \\
split_c(\infty)
           &= 0 \bmax 0 \bmax split_c(\infty) + 0 = split_c(\infty)
\end{align*}
and
\begin{align*}
split_p(0) &= (0, 0) \\
split_p(1) &= (0, 0) \bmax (1, 0) \\
           &= (1, 0) \\
split_p(n) &= (0, 0) \bmax (1, 0) \bmax 
              (\smfst(\smsnd(split(n-2)))+1, \smsnd(\smsnd(split(n-2)))+1) \\
           &= (\smfst(split_p(n-2))+1, \smsnd(split_p(n-2))+1) \\
split_p(\infty)
           &= (\smfst(split_p(\infty))+1, \smsnd(split_p(\infty))+1) \\
\end{align*}

\begin{wrapfigure}{R}{3in}
\begin{callout}
Of course, this is the same solution that we would get by finding a least
upper bound with respect to the usual information order (partial function
containment) for finite~$n$.  Presumably that has something to do with the
fact that $F$ is also monotone in the information order and
the least upper bound with respect to the converse
size order or ordinary information order occurs at~$f_\omega$.
\end{callout}
\end{wrapfigure}

So how do we go about solving these recurrences?  Formally the solution is
the least upper bound of the iterates of the functional defined by the
recurrence.  But we should keep in mind that when we say ``least upper
bound,'' that is in the order~$\geq$ (i.e., the converse of the size order).
Let's do that for $split_p$.  Let $F$ be defined by
\begin{align*}
F\,f\,0 &= (0, 0) \\
F\,f\,1 &= (1, 0) \\
F\,f\,n &= (\smfst(f(n-2)) + 1, \smsnd(f(n-2))+1) \\
F\,f\,\infty &= (\smfst(f\,\infty), \smsnd(f\,\infty))
\end{align*}
and let $f_k$ be the iterates of $F$:
\begin{align*}
f_0(x) &= \infty \\
f_{k+1}(x) &= F\,f_k
\end{align*}
By induction on~$k$ we have that
\[
f_k(x) = 
\begin{cases}
(\ceil{x/2}, \floor{x/2}),&x<k \\
(\infty,\infty),&x\geq k.
\end{cases}
\]
We then observe that $split_p(x) = (\ceil{x/2}, \floor{x/2})$ is an upper
bound of all the~$f_k$ (with respect to $\geq$), and that if $f$ is an upper
bound of all the~$f_k$ 
with respect to~$\szgeq$, then $f(x)\leq split_p(x)$ and hence $split_p$
is the least upper bound.  Thus $split_p$ is in fact the solution to the
recurrence.

Figure~\ref{fig:merge-syn-rec} shows the syntactic recurrence extracted
from~$merge$, again simplifying using equations that are valid in the
current model.
Using notation analogous to that used for $split$, we have that 
\begin{align*}
merge_p(0, l) &= l \\
merge_p(k, 0) &= 0 \bmax k \\
              &= k \\
merge_p(k, l) &= l \bmax k \bmax (merge_p(k-1,l)+1 \bmax merge_p(k,l-1)+1)
\end{align*}
and
\begin{align*}
merge_c(0, l) &= 0 \\
merge_c(k, 0) &= 0 \bmax 0 \\
              &= 0 \\
merge_c(k, l) &= 0 \bmax 0 \bmax (merge_c(k-1,l)+1 \bmax merge_c(k,l-1)+1)
\end{align*}
where we define $\infty-1$ to be~$\infty$.  Using an argument similar to
the one we used for~$split_p$, we see that
$merge_p(k, l) = k+l$ and $merge_c(k, l) = k+l$.
Here we are making use of the fact that 
$\cint = \overbrace{\cunit+\dots+\cunit}^{\text{$K$ times}}$ for some
large~$K$ and $\cbool = \csum{\cunit}{\cunit}$ for interpreting
$\cifkw$ expressions.  We must compute
\[
\den[\env{\envbind{x,y}{\infty_{\den{\cint}},\infty_{\den{\cint}}}}]
    {\cif{x'\leq y'}{e_0}{e_1}}
\]
where ${\cif{x'\leq y'}{e_0}{e_1}}$ is notation for
$\ccase{x'\leq y'}{\cwild}{e_0}{\cwild}{e_1}$.
Keeping in mind how we interpret sums, $\infty_{\den{\cint}}
= \den\cunit\disjunion\dotsb\disjunion\den\cunit$.
Furthermore, $x'\leq y'$ is itself a $\ccasekw$ expression with~$K^2$
branches.  Since we take the maximum (union, in this case) over all the
branches corresponding to values in $\infty_{\den\cint}$, we end up taking
that maximum over $\set{\sminj 0 *, \sminj 1 *}$, and this in turn tells us
that 
\[
\den[\env{\envbind{x,y}{\infty_{\den{\cint}},\infty_{\den{\cint}}}}]
    {\cif{x'\leq y'}{e_0}{e_1}} =
\den{e_0} \bmax \den{e_1}
\]
(which in this case does not depend on~$x'$ or~$y'$ because
$\den{\ccons{\cwild}{e}} = 1 + \den{e}$).

Finally, the syntactic recurrence for $msort$ is given in
Figure~\ref{fig:msort-syn-rec}.  Keep in mind that in
Figure~\ref{fig:merge-sort}, the ``identifiers'' $split$ and $merge$ in the
definition of $msort$ really stand for
$\sfun{split}{xs}{\cdots}$ and $\sfun{merge}{xsys}{\cdots}$, respectively.
Thus the use of $split$ and $merge$ in 
Figure~\ref{fig:msort-syn-rec} really stand for
$\cfix{split}{\cdots}$ and $\cfix{merge}{\cdots}$ from
Figures~\ref{fig:split-syn-rec} and~\ref{fig:merge-syn-rec}, respectively.
The key point is that the denotation of, e.g., $\ccost*{\capp{split}{zs}}$
is again $split_c(\den{zs})$, as we would expect.

The denotation of $msort_p$ is
\begin{align*}
msort_p(0) &= 0 \\
msort_p(1) &= 0 \bmax 1 = 1 \\
msort_p(n) &= 0 \bmax 1 \bmax merge_p(msort_p(k), msort_p(l))
              & & (k, l) = split_p(n) \\
           &= 0 \bmax 1 \bmax 
              merge_p(msort_p(\ceil{n/2}), msort_p(\floor{n/2})) \\
           &= 1 \bmax (msort_p(\ceil{n/2})+msort_p(\floor{n/2})).
\end{align*}
The denotation of $msort_c$ is
\begin{align*}
msort_c(0) &= 0 \\
msort_c(1) &= 0 \bmax 0 = 0 \\
msort_c(n) &= 0 \bmax 0 \bmax 
              split_c(n) + merge_c(k, l) + msort_c(k) + msort_c(l)
              & & (k, l) = split_p(n) \\
           &= merge_c(\ceil{n/2}, \floor{n/2}) +
              msort_c(\ceil{n/2}) + msort_c(\floor{n/2}) \\
           &= \ceil{n/2} + \floor{n/2} +
              msort_c(\ceil{n/2}) + msort_c(\floor{n/2}) \\
           &= n + msort_c(\ceil{n/2}) + msort_c(\floor{n/2})
\end{align*}
and the standard proof tells us that
$msort_c(n) \in O(n\lg n)$.

\section{Adapting the constructor counting model for quick sort}
\label{sec:quick-sort}

\begin{figure}
\[
\begin{array}{l}
\sfun*{part}{\spair x {xs}}
{\scaselist*{xs}
            {\spair\snil\snil}
            {y}{ys}
            {\slet*{\spair{ws}{zs}}
                   {\sapp{part}{\spair x {ys}}}
                   {\sif{\stick*{x \leq y}}
                        {\spair{ws}{\scons y {zs}}}
                        {\spair{\scons y {ws}}{zs}}
                   }
            }
}
\\ \\
\sfun{app}{\spair{xs}{ys}}
{\scaselist{xs}
           {ys}
           {x'}{xs'}
           {\scons{x'}{\sapp{app}{\spair{xs'}{ys}}}}
}
\\ \\
\sfun*{qsort}{xs}
{\scaselist*{xs}
            {\snil}
            {y}{ys}
            {\slet*{\spair{ws}{zs}}
                   {\sapp{part}{\spair{y}{ys}}}
                   {\slet*{\spair{ws'}{zs'}}
                          {\spair{\sapp{qsort}{ws}}{\sapp{qsort}{zs}}}
                          {\sapp{app}{\spair{ws'}{\scons{y}{zs'}}}}
                   }
            }
}
\end{array}
\]
\caption{Quick sort.}
\label{fig:quick-sort}
\end{figure}

It would seem that the constructor counting model of the previous section
would be appropriate for analyzing any list algorithm in terms of the length
of the argument list.  However, that is not the case; let us see what the
problem is when analyzing the worst-case cost of deterministic quick sort as
given in Figure~\ref{fig:quick-sort}.  The trouble arises in the analysis of
$part$, for which the syntactic recurrence is given in
Figure~\ref{fig:qsort-syn-recs}.

\begin{figure}
\begin{align*}
& \cpot{\ctrans{\sfun{part}{(x,xs)}{\cdots}}} = \\
& \cfix*{part}{
    \clam{xxs}{
      \clet*{\cpair{x}{xs}}{xxs}{
      \ccaselist*
        {xs}
        {\czcost{\cpair{\cnil}{\cnil}}}
        {y}{ys}
        {
          \plusc{(\capp{part}{\cpair{x}{ys}})_c}{
            \clet*{\cpair{ws}{zs}}{(\capp{part}{\cpair{x}{ys}})_p}{
              \plusc*{1}{
                 \cif*{x\leq y}{
                   \czcost{\cpair{ws}{1+zs}}
                 }{
                   \czcost{\cpair{1+ws}{zs}}
                 }
              }
            }
          }
        }
      }
    }
  } \\
\\
& \cpot{\ctrans{\sfun{app}{xsys}{\cdots}}} = \\
& \cfix*{app}{
    \clam{xsys}{
      \clet*{\cpair{xs}{ys}}{xsys}{
        \ccaselist
          {xs}
          {\czcost{ys}}
          {x'}{xs'}
          {
            \czcost{1+\capp{app}{\cpair{xs'}{ys}}}
          }
      }
    }
  } \\
\\
& \cpot{\ctrans{\sfun{qsort}{xs}{\cdots}}} = \\
& \cfix*{qsort}{
    \clam{xs}{
      \ccaselist*
        {xs}
        {\czcost{\cnil}}
        {y}{ys}
        {
          \plusc*{\ccost*{\capp{part}{\cpair{y}{ys}}}}{
            \clet*{\cpair{ws}{zs}}{\cpot*{\capp{part}{\cpair{y}{ys}}}}{
              \plusc*{\ccost*{\capp{qsort}{ws}}+\ccost*{\capp{qsort}{zs}}}{
                \clet*{
                  \cpair{ws'}{zs'}
                }{
                  \cpair{\cpot*{\capp{qsort}{ws}}}{\cpot*{\capp{qsort}{zs}}}
                }{
                  \czcost{\capp{app}{\cpair{ws'}{1+zs'}}}
                }
              }
            }
          }
        }
    }
  }
\end{align*}
\caption{The syntactic recurrences for quick sort.}
\label{fig:qsort-syn-recs}
\end{figure}

As usual, let $part_p(X, x)$ be the semantic recurrence
$\den{\ctrans{\cpot*{\sfun{part}{xxs}{\cdots}}}}$.  Then the usual analysis
tells us that $part_p(X, x) = (x, x)$.  That is, the extracted recurrence
tells us that if $xs$ is a list of length~$n$, $part(x, xs)$ partitions~$xs$
into two lists, each of which has length $\leq n$.  Though a correct
statement (necessarily, since the model satisfies the size order rules), it
is too weak, because the semantic recurrences extracted for $qsort$ will then
tell us only that 
\[
  qsort_c(n) = part_c(\den{\cint}, n) + 2qsort_c(n-1),
\]
from which we will conclude that the cost of quick sort is $O(2^n)$.

The proximate problem is that the conditional in the definition of~$part$
is interpreted as the maximum of its branches, and the maximum in a product
is interpreted componentwise; in this case, the maximum potential of the
branches is $(1+ws, 1+zs)$.  One perspective of the underlying problem is
that our interpretation of products forces us to come up with a pair that
simultaneously bounds both branches.  While there is such a bound, it
obscures any non-trivial relationship between the components of the pairs
being bounded.  This is similar to the problem we
discussed with interpreting sums as the ordinary disjoint sum with an
additional top element; when we happen to have elements on both sides of the
sum, our only choice is to bound them by the top element.

\begin{figure}
\begin{align*}
\den{\cprod{\sigma_0}{\sigma_1}} 
  &= \powerset(\den{\sigma_0}\cross\den{\sigma_1}) \\
X \szleq Y
  &\Iff \forall(x,x')\in X\exists (y,y')\in Y.x\szleq y\And x'\szleq y' \\
\\
\den[\eta]{\cpair{e_0}{e_1}}
  &= \set{(\den[\eta]{e_0}, \den[\eta]{e_1})} \\
\den[\eta]{\clet{\cpair{x_0}{x_1}}{e'}{e}}
  &= \bigszlub\setst{\den[\envext\eta{x_0,x_1}{a_0,a_1}]{e}}
                    {(a_0,a_1)\in\den[\eta]{e'}} \\
\\
\smcsize_{\cprod{F_0}{F_1}}(X)
  &= \bigszlub\setst{\smcsize_{F_0}(a_0) + \smcsize_{F_1}(a_1)}
                    {(a_0, a_1)\in X}
\end{align*}
\caption{Interpreting products as powersets.}
\label{fig:csize-product-powerset}
\end{figure}

The problem is analogous to the one for sums, and the solution is likewise
similar:  we adapt the model by interpreting products as (a subset of) the
powerset rather than the ordinary cartesian product.  That is, we replace
the clauses for the interpretation of products from 
Figures~\ref{fig:csize-model-types}, \ref{fig:csize-model-terms},
and~\ref{fig:csize-model-csize-fn} with those in
Figure~\ref{fig:csize-product-powerset}.  Although the proofs are more
painful, the equations of Figure~\ref{fig:cc-model-eqns} are still valid in
the adapted model.  The additional pain in many cases arises from the
one-step unfolding of $\clist\sigma$.  We now have that
$\den[\env{\envbind\alpha\Ninfty}]{\csum\cunit{\cprod\sigma\alpha}} =
 \powerset(\set{*} \disjunion \powerset(\den\sigma\cross\Ninfty))$, and
hence a typical element of the unfolding has the form 
$\emptyset\disjunion X$ or $\set{*}\disjunion X$, where
$X\subseteq\powerset(\den\sigma\cross\Ninfty)$.  This makes computing
the denotation of $\ccasekw$ expressions unpleasant, but using
the fact that $\den{\cunfold n} =
\set{*}\disjunion\powerset(\den{\sigma}\disjunion\set{0,\dots,n-1})$, one finds
that it is the same as before:
\[
\den[\eta]{\ccaselist n {e_\cnil} {x}{xs} {e_\cconskw}} =
\den[\eta]{e_\cnil} \bmax
\den[\envext\eta{x,xs}{\infty,n-1}]{e_\cconskw}.
\]

This in turn allows us to compute the functions~$part_p$
and~$part_c$:
\begin{align*}
part_p(X) &=
\bigunion_{(a, n)\in X}\set[bigg]{
  \set{(0, 0)} 
  \union
  \bigunion_{(a, n)\in X, n>0}\set[Big]{
    \bigunion_{ (k, \ell) \in part_p\set{(a, n-1)} }\set[big]{
      \set{(k, 1+\ell), (1+k, \ell)}
    }
  }
} \\
part_c(X) &=
\bigszlub_{(a, n)\in X}\set[Big]{
  0
  \bmax
  \bigszlub_{(a, n)\in X, n>0}\set[big]{
    part_c\set{(a, n-1)}
  }
}
\end{align*}
Let us focus on~$part_p$.  If we let~$P$ be the functional defined by the
recurrence above and $f_r$ the iterates of~$P$ starting at
$f_0(X) = \powerset(\Ninfty\cross\Ninfty)$, then a bit of computation shows
us that for~$r\geq 1$,
\[
f_r(X) =
\begin{cases}
\emptyset,& X=\emptyset \\
\setst{(k,\ell)}{k+l\leq n},&
  X\not=\emptyset, (a, s)\in X\Implies s<r,
  n=\max\setst{s}{\exists a.(a, s)\in X} \\
\powerset(\Ninfty\cross\Ninfty),&
  X\not=\emptyset, \exists(a, s)\in X, s\geq r.
\end{cases}
\]
Thus $part_p(X)$ is given by
\[
part_p(X) =
\begin{cases}
\emptyset,&X=\emptyset \\
\setst{(k,\ell)}{k+l\leq n},&
  X\not=\emptyset,
  n=\max\setst{s}{\exists a.(a, s)\in X} < \infty \\
\powerset(\Ninfty\cross\Ninfty),&
  X\not=\emptyset,
  \max\setst{s}{\exists a.(a, s)\in X} = \infty
\end{cases}
\]
and similarly
\[
part_c(X) =
\begin{cases}
0,&X=\emptyset \\
n,&
  X\not=\emptyset,
  n=\max\setst{s}{\exists a.(a, s)\in X} < \infty \\
\infty,&
  X\not=\emptyset,
  \max\setst{s}{\exists a.(a, s)\in X} = \infty.
\end{cases}
\]

Skipping a few steps, let us now consider $qsort_c(n)$.  When $n>0$, our
current interpretation of products results in
\begin{align*}
qsort_c(n) 
  &= part_c(n-1) + \bigszlub\setst{qsort_c(k)+qsort_c(\ell)}
                                  {(k, \ell)\in part_p(y, n-1)} \\
  &= part_c(n-1) + \bigszlub\setst{qsort_c(k)+qsort_c(\ell)}
                                  {k+\ell\leq n-1} \\
  &= part_c(n-1) + \bigszlub\setst{qsort_c(k)+qsort_c(\ell)}
                                  {k+\ell = n-1}
\end{align*}
The last equality follows from the fact that $qsort_c$ is monotone, and we
observe that this is precisely the recurrence that the informal analysis
yields.

\section{Some closing thoughts}
\label{sec:closing-thoughts}

The reader should be aware that the musings in this section have not been
carefully thought through.  They might be meaningless or trivial on their
face, and certainly expose lacunae in the author's knowledge that he wishes
did not exist.

\subsection{Different axioms}

The recurrence language of \citet{kavvos-et-al:popl2020} does not include
the axiom $0\leq e$, which we have relied on here (in the argument that for
an incomplete derivation of the form $\ieval e 0$, $0\leq\ctrans e_c$).
Does this matter?  We lose the exact cost model of
\citet{danner-licata:jfp22}, in which $\szleq$ is interpreted as the
equality relation.  And it seems strange to not have an exact cost model.
But what about the standard full type structure in which $\C$ is interpreted
as~$\N^\infty$ with the usual order, and $\szleq_\sigma$ is the identity for
all other $\sigma$?  It seems like this sort of structure should be a model,
and I suspect you still get exact costs, because same size means same
(semantic) value.  But then in order to do the ``standard thing,'' and
ensure that types are interpreted by domains (at least with respect to the
information order) so that we can use the standard interpretation for
inductive types, either the information order is the standard one, in which
case it is not quite $\szgeq$, or the size order isn't quite the identity.
Either way, I think there is a little bit of work to be done here.  And
unfortunately what I've learned thinking about it is that I don't actually
understand the standard interpretation of inductive types as well as I
thought I did.

\citet{kavvos-et-al:popl2020} also includes a fixpoint induction type of
axiom and axioms that say that partial approximants of a fixpoint go down in
the size order.  It seems like the latter ought to correspond roughly to the
$\beta_{\cfixkw}$ axiom that we have here, though maybe only in models.  And
we might only be able to usefully compute bounds from extracted recurrences
in the semantics provided the model actually satisfies some form of fixpoint
induction.  The reason is that the recurrence
extracted from a recursive function definition is itself a fixpoint, and
hence its interpretation is defined as a limit of the finite approximants.
That sequence of approximants goes down in the size order.
To conclude useful information about the fixpoint itself, we need to know
that a lower bound on all the approximants is also a lower bound on the
limit.  So even if we don't need fixpoint induction in the syntax, we do
seem to need it in the semantics.  This doesn't seem to be an issue in the
constructor counting model here, but that might be because 
\infruleref{beta-fix} is actually an equality in this model.  So are there
interesting models in which it isn't?  Is this really even the operative
issue?

\subsection{Different cost models}

We keep talking about how this extraction technique should extend easily to
other cost models.  It is time to make good on that promise.  
\citet{raymond:thesis} approaches parallel complexity.
There, the cost annotations in the operational semantics describe work/span
cost graphs \citep{blelloch-greiner:fpca95}, 
and because the cost type in the recurrence language is
transparent, this is directly carried over into the recurrence language as
well.  Is there a way to have a more generic description of cost in the
source language so that the operational semantics doesn't have to be changed
for each cost model?  And what do we have to do to the now-opaque cost type
in the recurrence language so that it can be usefully interpreted as
work/span cost graphs?  I also think it would be especially nice if there
were models that resulted in probabilistic analyses.  What would such a
model be?  I'm thinking about the probabilistic analysis of deterministic
quick sort:  if all lists of a given length are equally likely, then the
expected cost is $O(n\lg n)$.  That sounds like 
$\den{\clist\sigma}$ should be a probability distribution for each~$n$, in
which case $qsort_c$ maps probability distributions to expected costs, so
$\C$ is interpreted as some kind of expectation.  Or perhaps we really need
to go with the idea that $\ccpy\sigma$ is really an algebra, and the cost
component potentially differs for different $\sigma$.

\subsection{Abstract types}

Something I've had a bee in my bonnet about for awhile is abstract types.
Consider heap sort, in which the elements of a list~$xs$ are entered into a
priority queue, then the priority queue is emptied out.  Assuming the cost
of the priority queue operations are $O(\lg n)$, the cost of the sorting
algorithm is $O(n\lg n)$.  The cost analysis needs to know nothing more
about the implementation of the priority queue beyond the cost of the
operations, any more than proving correctness needs to know anything more
about the implementation beyond its correctness.  So how can we bring this
sort of reasoning into this recurrence extraction setting?  Presumably we
want to use existential types to model abstract types, though for a first
pass I tend to lean toward the abstract type declarations
of \citet{mitchell-plotkin:popl85} for concreteness.  I think the idea is to
have both existential/abstract types in both the source and recurrence
language.  Of course, it is the semantics of the abstract type declarations
that is interesting.  I think what happens is that the abstract type can be
interpreted one way, with the concrete implementation interpreted another.
All that should be necessary is that there be the standard Galois connection
corresponding to the $\beta$ axioms in the recurrence language.  For
example, we might implement priority queues with leftist heaps.  The size of
a priority queue (i.e., the denotation of the abstract type) would just be
its size.  The denotation of the concrete type (leftist heaps) would have to
include rank (length of the rightmost branch), because that is the
quantity that the cost recurrences are defined in terms of.  On the side,
one proves that the rank is logarithmic in the size, so in fact the
denotation of the concrete type is probably includes size and rank---i.e.,
$\Ninfty\cross\Ninfty$, or possibly something like the pairs $(n, r)$ with
$r\leq \lg n$ (I forget the exact relation).  So what is the relation
between the abstract and concrete sizes?  I would guess that in fact we
interpret concrete sizes by $\powerset(\Ninfty\cross\Ninfty)$, the abstract
size~$n$ is mapped to $\setst{(n, r)}{r\leq \lg n}$, and the concrete
size~$X$ is mapped to $\bigszlub\setst{n}{(n, r)\in X}$.  The key point is
that the cost of the concrete leftist heap operations are linear in (concrete)
rank.  But that can then be translated to (concrete) size, which in turn
tells us that the cost of the abstract priority queue operations are
logarithmic in (abstract) size.  And from there, we can analyze the heap
sort algorithm itself.

In fact, it should be that we can go further.  It ought to be possible to
include some form of cost information in the abstract/existential type in
the source language (and hence in the recurrence language).
\citet{niu-et-al:popl22} tells us that \citet{acar-blelloch:Algorithms}
define a notion of \emph{cost signature} to go along with functional
signatures for abstract types, but a quick skim of the latter didn't turn it
up.  Nonetheless, it seems to consist of annotation functions with cost
information.  That cost information could presumably be used by the
$\stick{}$ operation as well as in recurrence extraction.  It would probably
also be carried over into the recurrence language.  A model of the
recurrence language would then have to validate those cost annotations.

\subsection{Things I don't know enough about}
\label{sec:dont-know}

We've proved soundness for recursive types, but haven't used them at all.
Part of that is because I don't understand models for arbitrary recursive
types.  But maybe that's a non-issue here; after all, the point is that we
would probably model recursive types by a simple notion of size.  Another
part is that I don't really know much about recursive types, and in
particular don't know much about typical algorithms that use them and how we
talk about their cost.

It seems like Proposition~\ref{prop:semantic-adequacy} should be provable
without the sensible ticking hypothesis, just as in
\citet{kavvos-et-al:popl2020}.  Is that true?  And is it interesting?  That
is, do we really care about the cost of a program that is not sensibly
ticked?  On the other hand, we haven't actually proved that our ticking of
merge sort and quick sort is sensible.  So this still takes some cleaning
up.

\appendix

\section{Syntactic recurrence simplifications}

\subsection{Merge sort}
\label{app:merge-sort-syn-rec-derivations}

Simplifications of the syntactic recurrences for the functions used to
define merge sort are given in Figures~\ref{fig:split-syn-rec},
\ref{fig:merge-syn-rec}, and~\ref{fig:msort-syn-rec}.

\begin{figure}
{\small
\begin{multline*}
\cpot{\ctrans{\sfun{split}{xs}{\cdots}}} \\
\begin{aligned}
&= \cfix*{split}{
     \clam{xs}{
       \ccaselist*
         {xs}
         {\ctrans{\spair{\snil}{\snil}}}
         {y}{ys}
         {\ctrans{\scaselist{ys}{\dotsc}{z}{zs}{\dotsc}}}
     }
   } \\
&= \cfix*{split}{
     \clam{xs}{
       \ccaselist*
       {xs}
       {
         \cbind 
           {\cpair{p_0}{p_1}} 
           {\cpair{\ctrans{\snil}}{\ctrans{\snil}}}
           {\cval{\cpair{p_0}{p_1}}}
       }
       {y}{ys}
       {
         \ccaselist*
           {ys}
           {
             \cbind*
               {\cpair{p_0}{p_1}}
               {\cpair{\ctrans{\scons{y}{\snil}}}{\ctrans\snil}}
               {\cval{\cpair{p_0}{p_1}}}
           }
           {z}{zs}
           {
             \cbind*
               p
               {\ctrans{\sapp{split}{zs}}}
               {
                 \clet*{\cpair{as}{bs}}{p}{
                   \ctrans{
                     \spair{\scons{y}{as}}{\scons{z}{bs}}
                   }
                 }
               }
           }
       }
     }
   } \\
&= \cfix*{split}{
     \clam{xs}{
       \ccaselist^
       {xs}
       {
         \cbind 
           {\cpair{p_0}{p_1}} 
           {\cpair{\withcost 0 0}{\withcost 0 0}}
           {\cval{\cpair{p_0}{p_1}}}
       }
       {y}{ys}
       {
         \ccaselist^
           {ys}
           {
             \cbind*
               {\cpair{p_0}{p_1}}
               {\cpair{\czcost*{1+\cnil}}{\withcost 0 0}}
               {\cval{\cpair{p_0}{p_1}}}
           }
           {z}{zs}
           {
             \cbind*
               p
               {\capp{split}{zs}}
               {
                 \clet*{\cpair{as}{bs}}{p}{
                   \cbind*
                     {\cpair{q}{q'}}
                     {\cpair{\czcost*{1+as}}{\czcost*{1+bs}}}
                     {\cval{\cpair{q}{q'}}}
                 }
               }
           }
       }
     }
   } \\
&= \cfix*{split}{
     \clam{xs}{
       \ccaselist*
       {xs}
       {
         \czcost{\cpair 0 0}
       }
       {y}{ys}
       {
         \ccaselist*
           {ys}
           {
             \czcost{\cpair 1 0}
           }
           {z}{zs}
           {
             \plusc*
               {\ccost*{\capp{split}{zs}}}
               {
                 \clet*{\cpair{as}{bs}}{\cpot*{\capp{split}{zs}}}{
                   \czcost{\cpair{1+as}{1+bs}}
                 }
               }
           }
       }
     }
   } \\
\end{aligned}
\end{multline*}
}
\caption{The syntactic recurrence for $split$.}
\label{fig:split-syn-rec}
\end{figure}

\begin{figure}
{\small
\begin{multline*}
\cpot{\ctrans{\sfun{merge}{xsys}{\cdots}}} \\
\begin{aligned}
&= \cfix*{merge}{
     \clam{xsys}{
       \cbind*{p}{\cval xsys}{
         \clet*{\cpair{xs}{ys}}{p}{
           \ccaselist*
             {xs}
             {\cval ys}
             {x'}{xs'}
             {
               \ccaselist^
                 {ys}
                 {\cval{xs}}
                 {y'}{ys'}
                 {
                   \ctrans{\sif\cdots\cdots\cdots}
                 }
             }
         }
       }
     }
   } \\
&= \cfix*{merge}{
     \clam{xsys}{
       \clet*{\cpair{xs}{ys}}{xsys}{
         \ccaselist*
           {xs}
           {\czcost{ys}}
           {x'}{xs'}
           {
             \ccaselist^
               {ys}
               {\czcost{xs}}
               {y'}{ys'}
               {
                 \cbind*{p}{\ctrans{\stick{x'\leq y'}}}{
                   \cif*
                     p
                     {\ctrans{\scons{x'}{\sapp{merge}{\spair{xs'}{ys}}}}}
                     {\ctrans{\scons{y'}{\sapp{merge}{\spair{xs}{ys'}}}}}
                 }
               }
           }
       }
     }
   } \\
&= \cfix*{merge}{
     \clam{xsys}{
       \clet*{\cpair{xs}{ys}}{xsys}{
         \ccaselist*
           {xs}
           {\czcost{ys}}
           {x'}{xs'}
           {
             \ccaselist*
               {ys}
               {\czcost{xs}}
               {y'}{ys'}
               {
                 \plusc*
                   1
                   {
                     \cif*
                       {x'\leq y'}
                       {
                         \clet*{\chi}{\capp{merge}{\cpair{xs'}{ys}}}{
                           \withcost
                             {1 + {\cpot\chi}}
                             {\ccost\chi}
                         }
                       }
                       {
                         \clet*{\chi}{\capp{merge}{\cpair{xs}{ys'}}}{
                           \withcost
                             {1 + {\cpot\chi}}
                             {\ccost\chi}
                         }
                       }
                   }
               }
           }
       }
     }
   }
\end{aligned}
\end{multline*}
}
\caption{The syntactic recurrence for $merge$.  We write
$\clet\chi{e'}{e}$ for $\subst{e}{e'}{\chi}$.}
\label{fig:merge-syn-rec}
\end{figure}

\begin{figure}
{\small
\begin{multline*}
\cpot{\ctrans{\sfun{msort}{xs}{\cdots}}} \\
\begin{aligned}
&= \cfix*{msort}{
     \clam{xs}{
       \ccaselist^
         {xs}
         {\cval\cnil}
         {y}{ys}
         {
           \ccaselist^
             {ys}
             {\czcost{\ccons{y}{\cnil}}}
             {\cwild}{\cwild}
             {
               \cbind*
                 p
                 {\ctrans{\sapp{(\sfun{split}{xs}{\cdots})}{xs}}}
                 {
                   \clet*{\cpair{ws}{zs}}{p}{
                     \ctrans{
                       \sapp
                         {(\sfun{merge}{xsys}{\cdots})}
                         {\cpair{\capp{msort}{ws}}{\capp{msort{zs}}}}
                     }
                   }
                 }
             }
         }
     }
   } \\
&= \cfix*{msort}{
     \clam{xs}{
       \ccaselist^
         {xs}
         {\czcost{0}}
         {y}{ys}
         {
           \ccaselist^
             {ys}
             {\czcost{1}}
             {\cwild}{\cwild}
             {
               \cbind*
                 p
                 {
                   \plusc*
                     {
                       \ccost{\ctrans{\sfun {split} {xs} {\cdots}}} +
                       \ccost{\ctrans{xs}}
                     }
                     {
                       \capp
                         {\cpot{\ctrans{\sfun {split} {xs} {\cdots}}}}
                         {\cpot{\ctrans{xs}}}
                     }
                 }
                 {
                   \clet*{\cpair{ws}{zs}}{p}{
                     \plusc*
                     {
                       \ccost{\ctrans{\sfun{merge}{xsys}{\cdots}}} +
                       \ccost{\ctrans{
                         \spair{\sapp{msort}{ws}}{\sapp{msort}{zs}}
                       }}
                     }
                     {
                       \capp
                         {\cpot{\ctrans{\sfun{merge}{xsys}{\cdots}}}}
                         {\cpot{\ctrans{
                           \spair{\sapp{msort}{ws}}{\sapp{msort}{zs}}
                         }}}
                     }
                   }
                 }
             }
         }
     }
   } \\
&= \cfix*{msort}{
     \clam{xs}{
       \ccaselist^
         {xs}
         {\czcost{0}}
         {y}{ys}
         {
           \ccaselist^
             {ys}
             {\czcost{1}}
             {\cwild}{\cwild}
             {
               \cbind*
                 p
                 {\capp{split}{xs}}
                 {
                   \clet*{\cpair{ws}{zs}}{p}{
                     \plusc*
                     {
                       \ccost*{\capp{msort}{ws}} +
                       \ccost*{\capp{msort}{zs}}
                     }
                     {
                       \capp
                         {merge}
                         {
                           \spair
                             {\cpot*{\capp{msort}{ws}}}
                             {\cpot*{\capp{msort}{zs}}}
                         }
                     }
                   }
                 }
             }
         }
     }
   } \\
&= \cfix*{msort}{
     \clam{xs}{
       \ccaselist^
         {xs}
         {\czcost{0}}
         {y}{ys}
         {
           \ccaselist^
             {ys}
             {\czcost{1}}
             {\cwild}{\cwild}
             {
               \plusc*
                 {\ccost*{\capp{split}{xs}}}
                 {
                   \clet*{\cpair{ws}{zs}}{\cpot*{\capp{split}{xs}}}{
                     \plusc*
                     {
                       \ccost*{\capp{msort}{ws}} +
                       \ccost*{\capp{msort}{zs}}
                     }
                     {
                       \capp
                         {merge}
                         {
                           \spair
                             {\cpot*{\capp{msort}{ws}}}
                             {\cpot*{\capp{msort}{zs}}}
                         }
                     }
                   }
                 }
             }
         }
     }
   } \\
\end{aligned}
\end{multline*}
}
\caption{The syntactic recurrence for $msort$.}
\label{fig:msort-syn-rec}
\end{figure}
